\documentclass[a4paper,onecolumn,11pt,accepted=2021-04-19]{quantumarticle}
\pdfoutput=1
\usepackage[utf8]{inputenc}
\usepackage[english]{babel}
\usepackage[T1]{fontenc}
\usepackage{amsmath}
\usepackage{hyperref}

\usepackage{tikz}
\usepackage{lipsum}
\usepackage{bm}
\newtheorem{theorem}{Theorem}

\usepackage[numbers,sort,compress]{natbib}
\newcommand{\nn}{\nonumber \\}

\usepackage[normalem]{ulem}

\usepackage{graphicx}
\usepackage{dcolumn}
\usepackage{bm}
\usepackage{braket}

\usepackage[bottom]{footmisc}
\def\>{\rangle}
\def\<{\langle}

\usepackage{amssymb}
\newtheorem{proposition}{Proposition}
\newtheorem{corollary}{Corollary}
\newtheorem{conjecture}{Conjecture}
\usepackage{braket}

\begin{document}

\title{Photonic quantum data locking}

\author{Zixin Huang}
\email{zixin.huang@sheffield.ac.uk}
\affiliation{ Department of Physics \& Astronomy, University of Sheffield, UK }

\author{Peter P. Rohde}
\affiliation{Centre for Quantum Software \& Information (QSI), Faculty of Engineering \& Information Technology
University of Technology Sydney, NSW 2007, Australia}

\author{Dominic W. Berry}

\affiliation{Department of Physics and Astronomy, Macquarie University, Sydney, New South Wales 2109, Australia}

\author{Pieter Kok}
\affiliation{ Department of Physics \& Astronomy, University of Sheffield, UK }

\author{Jonathan P. Dowling 
}
\affiliation{Hearne Institute for Theoretical Physics and Department of Physics \& Astronomy, Louisiana State University, Baton Rouge, Louisiana 70803, USA}

\affiliation{National Institute of Information and Communications Technology,
4-2-1, Nukui-Kitamachi, Koganei, Tokyo 184-8795, Japan}

\affiliation{NYU-ECNU Institute of Physics at NYU Shanghai, Shanghai 200062, China}
\affiliation{CAS-Alibaba Quantum Computing Laboratory, USTC, Shanghai 201315, China}

\author{Cosmo Lupo}\email{c.lupo@sheffield.ac.uk}
\affiliation{ Department of Physics \& Astronomy, University of Sheffield, UK }

\maketitle

\begin{abstract}
\footnote{This paper is dedicated to the memory of Professor Jonathan P. Dowling.}
 Quantum data locking is a quantum phenomenon that allows us to encrypt a long message with a small secret key with information-theoretic security. 
This is in sharp contrast with classical information theory where, according to Shannon, the secret key needs to be at least as long as the message. 
Here we explore photonic architectures for quantum data locking, where information is encoded in multi-photon states and processed using multi-mode linear optics and photo-detection, with the goal of extending an initial secret key into a longer one.
The secret key consumption depends on the number of modes and photons employed. In the no-collision limit, where the likelihood of photon bunching is suppressed, the key consumption is shown to be logarithmic in the dimensions of the system.
Our protocol can be viewed as an application of the physics of Boson Sampling to quantum cryptography.
Experimental realisations are challenging but feasible with state-of-the-art technology, as techniques recently used to demonstrate Boson Sampling can be adapted to our scheme (e.g., Phys.\ Rev.\ Lett.\ \textbf{123}, 250503, 2019).
\end{abstract}

\section{Introduction}
\noindent

In classical information theory, a celebrated result of Shannon states that a message of $N$ bits can only be encrypted using a secret key of at least $N$ bits \cite{shannon1949communication}. This result, which lays the foundation of the security of the 
one-time pad, does not necessarily apply when information is encoded into a quantum state of matter or light.

The phenomenon of Quantum Data Locking (QDL), first discovered by DiVincenzo \textit{et al.} \cite{DiVin}, shows that a message of $N$ bits, when encoded into a quantum system, can be encrypted with a secret key of $k \ll N$ bits.
QDL guarantees information-theoretic security against an adversary who is forced to measure their share of the quantum system as soon as they obtain it, for example because they do not have a quantum memory \cite{PRX}, or after a known time, for example if they have a quantum memory with limited storage time \cite{PRL,Entropy}.
\color{black}
In QDL, a secret key of $k$ bits is used to uniquely identify a code, i.e., a basis among a given set of $2^k$ orthogonal bases in a Hilbert space of dimensions $d$. The elements of said basis are then used to reliably encode $N$ bits of information. 
It is known that there exist choices of $k \ll N$ bases such that only a negligibly small amount of information, as quantified by the accessible information, will leak if one attempts to measure the quantum state without knowing the secret key \cite{CMP,Fawzi13}. 
From a physical point of view, this is related to the fact that the $2^k$ bases correspond to non-commuting observables \cite{Wehner_2010,Coles}.

Initial works on QDL focused on abstract protocols that required control over an asymptotically large Hilbert space, including, for example, the ability to sample random unitaries according to the Haar measure \cite{CMP,Fawzi13}, or to perform universal quantum computation \cite{Fawzi13}.
Some more recent works have instead explored practical protocols that only require moderate control over a Hilbert space of relatively small size \cite{lloyd2013quantum,PRX,Winter2017,PRL,NJP}. This has led to one of the first experimental demonstrations of QDL \cite{Lum} (see also Ref.~\cite{Pan2016}). In this experiment, information was encoded into the transverse wave-vector of an heralded photon, and an array of spatial light modulators was used to generate pseudo-random unitary transformations of the single-photon wave front.
A setup encoding information in time of arrival was discussed in Ref.\ \cite{Notaros}.

In this paper, we define and characterize a family of QDL protocols that encode information using $n$ photons scattered over $m$ optical modes. Examples include spatial modes, temporal modes, and angular momentum \cite{Cozzolino}.
The goal of these protocols is to extend an initial secret key into a longer one.
While encoding information using $1$ photon over \textcolor{black}{$m$} modes is more practical, it is also highly inefficient as it encodes no more than \textcolor{black}{$(\log{m})/m$} bits per mode. %
By contrast, using multiple photons we can encode up to $1$ bit per mode. 
As our QDL protocols are well defined also for relatively small values of $n$ and $m$, experimental demonstrations are feasible with current state-of-the-art technology. In particular, the same techniques used to demonstrate Boson Sampling can be adapted to multiphoton multimode QDL
\cite{PhysRevLett.118.190501, PhysRevLett.121.250505, PhysRevLett.123.250503}.

Extending the theoretical security analysis of QDL to multiphoton states presents some technical challenges that we address and solve here. 
These challenges are related to the peculiar properties of the group of linear optical passive (LOP) unitaries.
LOP unitaries define a representation of the group $U(m)$, which becomes irreducible in the subspaces with definite photon number $n$ \cite{OSID2006}.
Hence previous proof techniques that are based on the properties of the fundamental representation of $U(m)$ cannot be directly applied to characterise multiphoton QDL.


The structure of the paper follows.
In Sec.~\ref{sec:qdl} we introduce the formalism of QDL. 
In Sec.~\ref{Sec:Protocol} we describe our protocol.
Preliminary considerations are presented in Sec.~\ref{Sec:prel}, and our main results are in Sec.~\ref{sec:results}.
Mathematical proofs are discussed in Secs.~\ref{Sec:Proof1} and~\ref{Sec:col1}.
Noisy communication channels are addressed in Sec.~\ref{Sec:real}.
Section~\ref{Sec:conclusion} is for conclusions.
Technical results are given in the Appendices~\ref{Sec:Chernoff}-\ref{Sec:loss}.

\section{Quantum data locking}\label{sec:qdl}

A typical QDL protocol requires a sender (conventionally called Alice) and a receiver (Bob) that can communicate through a quantum channel. Each use of the communication channel allows Alice to transfer a quantum state that lives in a Hilbert space of dimensions $d$.

\begin{figure}[t!]\centering
\includegraphics[trim = 0.3cm 0cm 0cm 0cm, clip, width=0.6\linewidth]{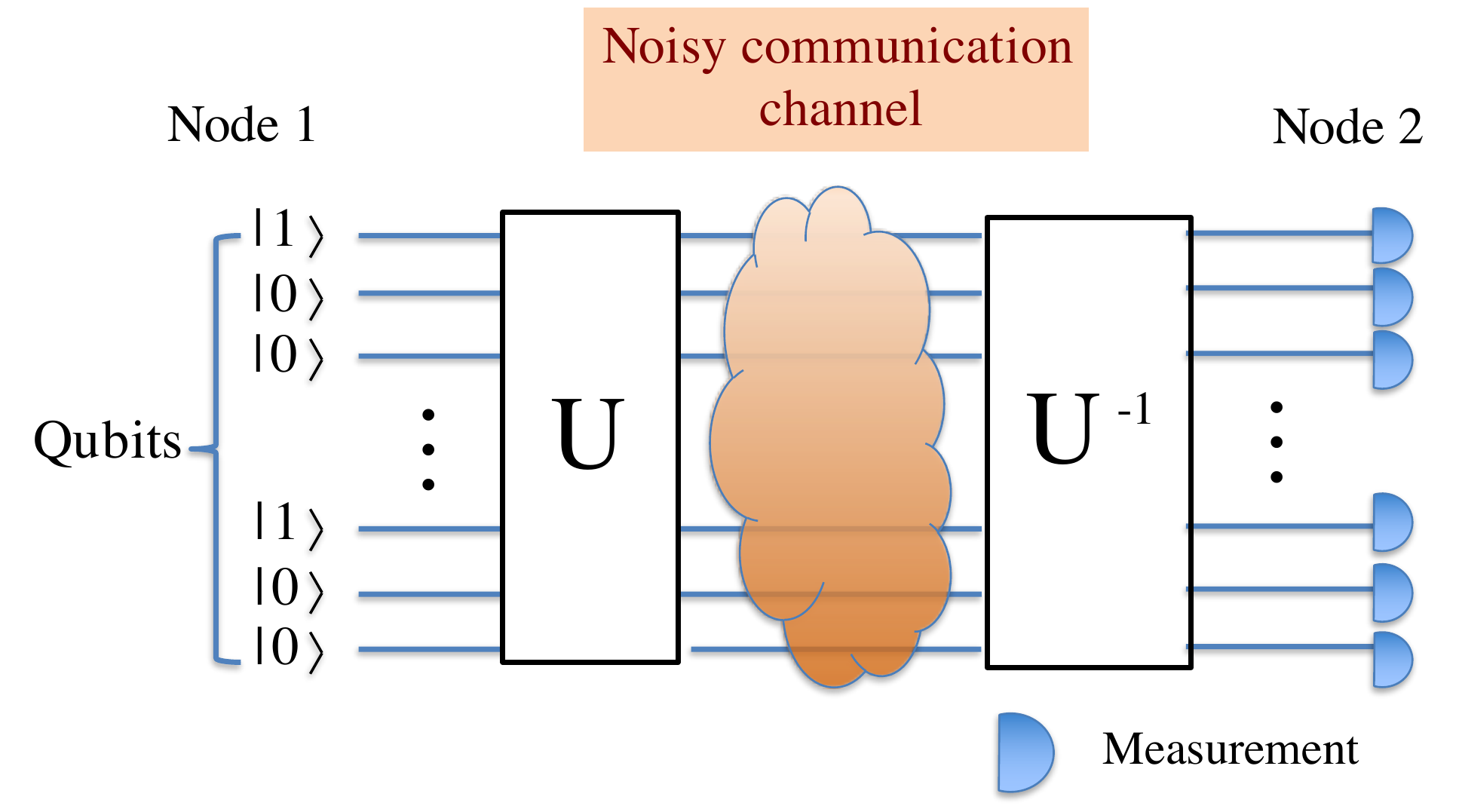} 
  \caption{\label{f:schematic} Circuit layout of the QDL protocol. 
  The code words $\ket{\psi_j}$ are chosen from the ${m \choose n}$ possible permutations of $n$ photons in $m$ modes. 
  Alice and Bob share a secret key in advance, which Alice uses to encrypt her photonic input message by applying a unitary $U$.
  Bob decrypt by applying the inverse operation $U^{-1}$.}
\end{figure}

The general form of a QDL protocol follows (Fig.~\ref{f:schematic}).
\begin{enumerate}
\item In advance, Alice and Bob publicly agree upon a set of $K$ unitary matrices in $U(d)$, say $\{ U_k \}_{k=1,\dots,K}$. 
Each unitary matrix identifies a basis in the Hilbert space, which is obtained by applying the unitary to the standard computational basis.
Denote as $\{ |j\rangle \}_{j=1,\dots,d}$ the elements of the computational basis. Then the unitary $U_k$ identifies the basis $\{ U_k|j\rangle \}_{j=1,\dots,d}$.
\item To send information to Bob, first Alice uses a secret key of $\log{K}$ bits to choose one particular unitary transformation, i.e., one particular basis in the agreed set of $K$ bases. 
\item Alice selects $M$ basis vectors, $ \{ U_k|j_x\rangle \}_{x=1,\dots,M}$ from the chosen basis and use them as a code to send $\log{M}$ bits of classical information through the quantum channel. 
%
%
%
%
%
%
%
%
This encoding of classical information into a quantum system $A$ is described by the classical-quantum state
\begin{align}\label{rhoXA}
    \rho_{XA}^k = \frac{1}{M} \sum_{x=1}^M |x\rangle_X \langle x| \otimes U_k |j_x\rangle_A \langle j_x | U_k^\dag \, ,
\end{align}
where $X$ is the classical variable encoded by Alice, which is represented by a set of $M$ orthogonal vectors $\{ |x\rangle \}_{x=1,\dots,M}$ in a dummy quantum system. 

\end{enumerate}

In this work we assume that different code words have equal probability. As the goal of the protocol is to extend an initial secret key into a longer one, using equally probable code words is a natural assumption.
It makes the analysis of the QDL protocol easier, although it can be relaxed \cite{Dupuis13,other}.

The code words prepared by Alice are then sent to Bob through a quantum channel described as a completely positive and trace preserving map $\mathcal{N}_{A \to B}$ that transforms Alice's system $A$ into Bob's system $B$. 
The channel maps the state in Eq.\ (\ref{rhoXA}) into
\begin{align}\label{rhoXBk}
    \rho_{XB}^k = \frac{1}{M} \sum_{x=1}^M |x\rangle_X \langle x| \otimes \mathcal{N}_{A \to B} \left( U_k |j_x\rangle_A \langle j_x | U_k^\dag \right) \, .
\end{align}

We ask a QDL protocol to have the properties of correctness and security.

\textit{Correctness.} The property of correctness requires that, if Bob knows the secret key used by Alice to chose the code words, then he is able to decode reliably.
For example, if the channel is noiseless, then $\mathcal{N}$ is the identity map and
\begin{align}\label{rhoXBk_0}
    \rho_{XB}^k = \frac{1}{M} \sum_{x=1}^M |x\rangle_X \langle x| \otimes U_k |j_x\rangle_A \langle j_x | U_k^\dag \, .
\end{align}
\color{black}
In this case, Bob can simply apply the inverse unitary, $U_k^{-1}$, followed by a measurement in the computational basis. In this way, Bob can decode with no error for any $M \leq d$.
If the channel is noisy, Alice and Bob can still communicate reliably at a certain rate of $r < \log{d}$ bits per channel use. This is possible by using error correction at any rate below the channel capacity, $r_\mathrm{max} = I(X;Y|K)$ \cite{wilde2013quantum}. Here $I(X;Y|K)$ denotes the mutual information between the input variable $X$ and the output of Bob's measurement $Y$, given the shared secret key $K$. 
Notice that here we need classical error correction and not quantum error correction, as the goal of Alice and Bob is to exchange classical information and not quantum information. 
Furthermore, we apply \textit{post facto} error correction, as it is commonly done in quantum key distribution \cite{postfacto}, in which error correcting information is sent independently on a classical authenticated public channel.

We emphasize the importance of the assumption that the adversary has no quantum memory for the security of post facto error correction. This assumption guarantees that a potential eavesdropper has already measured their share of the quantum system when the error correction information is exchanged on a public channel. If $b$ bits of error correcting information are communicated on a public channel, then the eavesdropper cannot learn more than $b$ bits of information about the message \footnote{To see that the public channel for error correction does not render the protocol insecure, we note that Eve's additional information about the secret key is bounded by classical information theory as follows. Let $X$ be the message sent by Alice, $Z$ the output of Eve's measurement, and $I(X;Z)$ the mutual information. After error correction, Eve obtains a bit string $C(X)$. Hence, we need to consider the mutual information $I(X ; Z C(X))$.
It follows from the property of \textit{incremental proportionality} \cite{DiVin} of the mutual information that $I(X ; Z C(X)) \leq I(X;Z) + H(C(X))$, where $H(C(X))$ is the entropy of $C(X)$. This implies that, knowing $C(X)$ after she measured the quantum system, Eve cannot learn more than $H(C(X))$ bits about the message $X$.}.
If instead the eavesdropper has a quantum memory with storage time $\tau$, then Alice and Bob need to wait for a time larger than $\tau$ after the quantum signal have been transmitted and before proceeding with post facto error correction. In this work we assume that Alice and Bob know an upper bound on $\tau$.
\color{black}

\textit{Security.} The property of security requires that, if Bob does not know the secret key, he can obtain no more than a negligibly small amount of information about Alice's input variable $X$.
To clarify this, consider that, if Bob does not know the secret key used by Alice, then his description of the classical quantum state is the average of Eq.\ (\ref{rhoXBk}),
\begin{align}\label{rhoXB}
    \rho_{XB} = \frac{1}{K} \sum_{k=1}^K \frac{1}{M} \sum_{x=1}^M |x\rangle_X \langle x| \otimes \mathcal{N}_{A \to B} \left( U_k |j_x\rangle_A \langle j_x | U_k^\dag \right) \, .
\end{align}
In QDL, the security is quantified using the accessible information \cite{DiVin,CMP} (or similar quantities \cite{Fawzi13,Dupuis13,Adamczak_2017}).
Recall that the accessible information $I_\mathrm{acc}(X;B)_\sigma$ is defined as the maximum information that Bob can obtain about $X$ by measuring his share of the state $\sigma$, that is,
\begin{align}
    I_\mathrm{acc}(X;B)_\sigma = \max_{M_{B \to Y}} I(X;Y) \, ,
\end{align}
where the optimization is over the measurement maps $M_{B \to Y}$ on system $B$, and $I(X;Y)$ is the mutual information between $X$ and the outcome $Y$ of the measurement.
The security property can be defined in different ways, depending on how the state $\sigma$ is chosen. Here we consider a strong notion of QDL \cite{PRX} and put 
\begin{align}\label{sigmaXB}
    \sigma_{XB} = \frac{1}{K} \sum_{k=1}^K \frac{1}{M} \sum_{x=1}^M |x\rangle_X \langle x| \otimes U_k |j_x\rangle_A \langle j_x | U_k^\dag \, .
\end{align}
\textcolor{black}{This is equivalent to saying that the information remains encrypted even if Bob is capable of accessing the quantum resource directly without the mediation of a noisy channel. 
The data processing inequality \cite{wilde2013quantum} then implies that the protocol is secure for noisy channels too.}
In conclusion, we say that the protocol is secure if $I_\mathrm{acc}(X;B) = O( \epsilon \log{M} )$, with $\epsilon$ arbitrarily small. This means that only a negligible fraction of the information can be obtained by measuring the quantum state without having knowledge of the secret key.

Intuitively, we expect that the larger $K$, the smaller the accessible information. This intuition has been proven true using tools from large deviation theory and coding theory \cite{CMP,Fawzi13,PRL}.
The mathematical characterization of a QDL protocol consists in obtaining, for given $\epsilon > 0$, an estimate of the minimum integer $K_\epsilon$ such that there exist choices of $K = K_\epsilon$ bases that guarantee $I_\mathrm{acc}(X;Y) = O( \epsilon \log{M} )$.

Finally, the net secret key rate that can be established between Alice and Bob, through a noisy communication channel $\mathcal{N}$, is 
\begin{align}
    r_\mathrm{QDL} = \beta I(X;Y|K) - \log{K_\epsilon} \, ,
\end{align}
where $\beta \in (0,1)$ is the efficiency of error correction, and we have subtracted the initial amount $\log{K_\epsilon}$ of secret bits shared between Alice and Bob. We emphasise that the mutual information $I(X;Y|K)$ depends on the particular noisy channel, whilst $\log{K_\epsilon}$ is universal.
The noisier the channel, the smaller $I(X;Y|K)$, which accounts for the error correction overhead. The factor $\beta$ accounts for the fact that practical error correction requires more overhead than expected in theory.

\section{Multiphoton encoding}\label{Sec:Protocol}

Let $n$ photons be sent into $m$ optical modes of an interferometer with at most one photon per input mode. The input modes $\hat{\bm{a}}$ evolve into $U\hat{\bm{a}}U^\dagger$, with $U$ the unitary transformation describing the interferometer:
\begin{align}
U a_i^\dagger U^\dagger = \sum_{j=1}^m U_{i,j} a_j^\dagger \, .
\end{align}

\color{black}
A passive multi-mode interferometer realises a unitary transformation that preserves the total photon number. The set of all possible transformations that can be realised in this way defines the group of linear passive optical (LOP) unitary transformations, which is isomorphic to the $m$-dimensional unitary group $U(m)$ (see e.g.\ Ref.\ \cite{OSID2006}).
By Shur's lemma, the group of LOP unitaries has irreducible representations in the subspaces with definite photon number.
For applications to photonic QDL, the representation with $1$ photon has been studied in previous works \cite{lloyd2013quantum,PRX}. This representation has the unique feature of being the fundamental representation of $U(m)$. 
However, representations with higher photon number that we are considering here are no longer the fundamental representation.

\color{black}

The output from the interferometer prior to photo-detection can be expanded in the photon-number basis:
\begin{align}\label{eq:t298wuhrs}
\ket{\psi_\text{out}} = \sum_{\bm{n}} \lambda_{\bm{n}}\ket{n_1 n_2  \dots n_{m}} \, ,
\end{align}
where $\bm{n} = (n_1, n_2 , \dots ,n_{m})$ denotes a photon-number configuration with $n_i$ photons in the $i$-th mode and $\lambda_{\bm{n}}$ its amplitude. 

The aim of this paper is to characterize a particular family of QDL protocols, where information is encoded into $m \geq 2$ optical modes using $n>1$ photons. We define the code words by putting photons on different modes, with no more than one photon per mode. 

In this way we obtain a code book $\mathcal{C}^m_n$ that contains $C = { m \choose n }$ code words, whereas the overall Hilbert space defined by $n$ photons on $m$ modes has dimensions $d = {n+m-1 \choose n}$ (this includes states with more than one photon in a given mode).
For example, with $m=4$ modes and $n=1$ photon, we have the $C=4$ code words $|1000\rangle$, $|0100\rangle$, $|0010\rangle$, $|0001\rangle$.
With $n=2$ photons, we instead obtain the $C=6$ code words $|1100\rangle$, $|0011\rangle$, $|1001\rangle$,  $|0110\rangle$, $|1010\rangle$, $|0101\rangle$.

The two users, Alice and Bob, are linked via an optical communication channel that allows Alice to send $m$ optical modes at the time.
Initially, we assume the channel is noiseless. 
Later we will extend to the case of a noisy channel.
The goal of the protocol, which is shown schematically in Fig.\ \ref{f:schematic}, is for Alice and Bob to expand an initial secret key of $\log K$ bits into a longer one.

For given $n$ and $m$, Alice defines a code book $\bar{\mathcal{C}}^m_n$ by choosing a subset of $M < C$ code words from $\mathcal{C}^m_n$. The code book is publicly announced. We denote the code words as $|\psi_x\rangle$, with $x = 1, \dots, M$.
To encrypt these code words, Alice applies an $m$-mode LOP unitary transformation from a set of $K$ elements $\{ U_k \}_{k=1,\dots,K}$. The unitary is determined by the value of her secret key of $\log{K}$ bits. 
We recall that any LOP unitary can be realised as a network of beam splitters and phase shifters \cite{Reck1994_PRL,Clements2017}.

We can directly verify the correctness property for a noiseless  communication channel. In this case, Bob, who knows the secret key, applies $U_k^{-1}$ and measures by photo-detection. He is then able to decrypt $\log{M}$ bits of information with no error. 
This implies that Alice and Bob can establish a key of $\log{M}$ bits for each round of the protocol. 

To characterise the secrecy of the QDL protocol, we need to identify the minimum key size $K_\epsilon$. This is the task that we accomplish in the following sections below.

\section{Preliminary considerations}\label{Sec:prel}

Before presenting our main results, we need to introduce some notation and preliminary results.
First, consider the following state,
\begin{equation}\label{barrhoE}
\bar{\rho}_B := \mathbb{E}_U[U | \psi \rangle \langle \psi | U^\dag] 
= \int dU \, U | \psi \rangle \langle \psi | U^\dag 
\, ,
\end{equation}
which is defined by taking the average over the LOP unitary $U$ acting on a state $\psi$.
Here $\mathbb{E}_U$ denotes the expectation value over the invariant measure (i.e., the Haar measure) on the group LOP unitary transformations acting on $m$ optical modes.
The choice of the invariant measure is somewhat arbitrary and other measures can be used, see e.g.\ Ref.\ \cite{PhysRevA.90.022326}.
In Eq.\ (\ref{barrhoE}), $\psi$ is a vector in the code book $\mathcal{C}^m_n$. By symmetry, $\bar{\rho}_B$ is independent of $\psi$.

\color{black}
Consider, as an example, the manifold of states with $n=3$ photons over $m=4$ modes. 
We denote as $\mathcal{H}_{(1,1,1,0)}$ the subspace spanned by vectors that have at most one photon in each mode, i.e., the linear span of $|1110\rangle$,  $|1101\rangle$, $|1011\rangle$, $|0111\rangle$.
The subspace $\mathcal{H}_{(1,1,1,0)}$ is characterised by a specific pattern describing how photons are distributed on the modes. 
Another subspace is $\mathcal{H}_{(2,1,0,0)}$, which is spanned by the vectors $|2100\rangle$, $|2010\rangle$..., etc. 
%
Finally, there exists one more photon pattern for $m=4$ modes and $n=3$ photons, denoted as $q=(3,0,0,0)$. The corresponding subspace $\mathcal{H}_{(3,0,0,0)}$ is spanned by the vectors $|3000\rangle$, $|0300\rangle$, $|0030\rangle$, $|0003\rangle$.
We label the different subspaces by $q\in\{(1,1,1,0),(2,1,0,0),(3,0,0,0) \}$.
These definitions naturally extend to any $n$ and $m$. In general, a patter of $n$ photons over $m$ modes is identified as $(n_1, n_2, \dots, n_m)$, for integers $n_j \geq n_{j+1}$ such that $\sum_{j=1}^m n_j = n$. We denote as $\mathcal{H}_q$ the corresponding subspace, and $P_q$ is the projector onto $\mathcal{H}_q$.
\color{black}


%
%

By symmetry, the state $\bar\rho_B$ is block-diagonal in the subspaces $\mathcal{H}_q$, i.e.,
\begin{align}\label{csum}
\bar\rho_B = \sum_q c_q P_q \, .
\end{align}
We are particularly interested in the smallest coefficient in this expansion,
\begin{align}
c_\mathrm{min} := \min_q c_q \, ,
\end{align}
which can be computed numerically for given $n$ and $m$. 
Examples are shown in Table \ref{Table:cmin}.
\color{black}
The results of our numerical estimations suggest that the minimum is always achieved for the pattern $q_\mathrm{min} = (1,1,1,..,0,0)$, i.e., when each mode contains at most $1$ photon.
An analytical expression for $c_{(1,1,1,..,0,0)}$ is given in Ref.~\cite{PhysRevA.94.042339},  
\begin{align}\label{c10}
c_{(1,1,1,..,0,0)} = {m+n-1 \choose n}^{-1} = \frac{1}{d} \, .
\end{align}

Supported by the results of our numerical search, we formulate the following conjecture:
\begin{conjecture}\label{con:cmin}
The smallest coefficient in the expansion in Eq.\ (\ref{csum}) is $c_\mathrm{min} = \min_q c_q = c_{(1,1,1,..,0,0)}$.
Equation (\ref{c10}) then implies 
\begin{align}
c_\mathrm{min} = {m+n-1 \choose n}^{-1} = \frac{1}{d} \, .
\end{align}
\end{conjecture}
We have used this conjecture to produce the plot in Fig.\ \ref{Fig:rates}. 
If the number of modes is much larger than the number of photons squared, $m \gg n^2 \gg 1$, the probability that two or more photons occupy a given mode is highly suppressed.
In this limit, we have 
$c_\mathrm{min} = n!/m^2$ (see Appendix \ref{Sec:speed}).
\color{black}

\begin{table}[ht]\center
\begin{tabular}{ |c|  c| c| c|}
\hline
$(m,n)$  & Photon occupancy pattern $q$ &  $c_q$    &  $n!/m^n$ 
\\ 
\hline\hline
$(6,2)$ & $(1, 1,0, 0,0,0)$  & $0.0471$ & $0.0556$ \\ 
        & $(2,0, 0,0,0,0)$   & $0.0977$ &          \\ 
\hline
$(10,2)$ & $(1, 1,0, 0,0,0,0,0,0,0)$  & $0.0181$ &  $0.02$ \\ 
         & $(2,0,0,0,0,0,0,0,0,0)$   &  $0.0369$   &     \\ 
\hline
$(20,2)$ & $(1,1,0,0, \dots)$    &  $0.00476$ & $0.005$ \\ 
         & $(2,0,0,0, \dots)$    & $0.00959$ &  \\
\hline
$(10,3)$ & $(1, 1, 1,0,0,0,0,0,0,0)$ & $0.00451$ & $0.006$ \\ 
         & $(1, 2,0,0,0,0,0,0,0,0)$   & $0.00914$  & \\
         & $(3,0,0,0,0,0,0,0,0,0)$    & $0.0283$  & \\ 
\hline
$(20,3)$ & $(1, 1, 1,0,\dots)$  & $0.000648$ &   $0.00075$ \\ 
         & $(1, 2,0,0,\dots)$  & $0.00131$ & \\
         & $(3,0,0,0,\dots)$ & $0.00398$ & \\
\hline
     %
     %
     %
     %
$(30,10)$  & $(1, 1, 1, 1,0,\dots)$ & $1.56\times 10^{-9}$ & $6.14 \times 10^{-9}$ \\ 
        &  $(1, 9,0,0,0,\dots)$     & $0.00068$  &  \\
        &  $(10,0,0,0,\dots)$        & $0.0072$ & \\ 
\hline
\end{tabular}
\caption{The table shows numerical estimates of the coefficient $c_q$ for several number of photon (n) and modes (m) and patterns of photon occupancy. The table also show the corresponding value of the coefficient $n!/m^n$, 
which is relevant in the regime where $m \gg n^2 \gg 1$.}\label{Table:cmin}
\end{table}

The other quantity we are interested in is 
\begin{align}\label{gamma0}
\gamma 
= \max_\phi \frac{ \mathbb{E}_U[|\langle \phi | U | \psi\rangle |^4] }
{\left( \mathbb{E}_U[|\langle \phi | U | \psi\rangle |^2] \right)^2 } \, ,
\end{align} 
where the maximum is over a generic $n$-photon vector $\phi$, and $\psi$ is a vector in the code book $\mathcal{C}^m_n$. Again, because of symmetry, $\gamma$ is independent of $\psi$.
Note that $\gamma$ quantifies how much the transition probability $|\langle \phi | U | \psi\rangle |^2$ changes when a random unitary is applied.
In the regime of $m \gg n^2 \gg 1$, an analytical bound can be computed and we obtain $\gamma \leq 2(n+1)$. This is discussed in Appendix \ref{Sec:speed}.

For generic values of $n$ and $m$, we estimate $\gamma$ numerically. To do that, we first expand the generic state $\phi$ as 
\begin{equation}
|\phi\rangle = \sum_{q,t} \alpha_{q,t} |\phi_{q,t}\rangle \, , 
\end{equation}
where $q$ identifies a subspace with given photon occupancy pattern, and $t$ labels the basis vectors within.
By symmetry, the following identities hold:
\begin{align}
\mathbb{E}_U \left[ \langle \phi_{q,t} | U | \psi \rangle \langle \psi| U^\dag |\phi_{q',t'} \rangle \right] 
= \delta_{qq'}\delta_{tt'} 
\mathbb{E}_U \left[ |\langle \phi_{q,t} | U | \psi \rangle |^2 \right] \, ,
\end{align}
and
\begin{align}
& \mathbb{E}_U \left[ \langle \phi_{q,t} | U | \psi \rangle \langle \psi| U^\dag |\phi_{q',t'} \rangle \langle \phi_{q'',t''} | U | \psi \rangle \langle \psi| U^\dag |\phi_{q''',t'''} \rangle\right] =  \nn
& \frac{\delta_{qq'}\delta_{tt'} \delta_{q''q'''}\delta_{t''t'''} + \delta_{qq'''}\delta_{tt'''} \delta_{q'q''}\delta_{t't''} }{2 \delta_{qq''}\delta_{tt''}} \, \mathbb{E}_U \left[ |\langle \phi_{q,t} | U | \psi \rangle|^2  |\langle \phi_{q'',t''} | U | \psi \rangle|^2 \right] \, .
\end{align}
This implies
\begin{align}
\mathbb{E}_U[|\langle \phi | U | \psi\rangle |^2]
= \sum_{q,t} | \lambda_{q,t} |^2 \, \mathbb{E}_U[|\langle \phi_{q,t} | U | \psi\rangle |^2] \, ,
\end{align}
and
\begin{align}
\mathbb{E}_U[|\langle \phi | U | \psi\rangle |^4]
&=  \sum_{q,t} |\lambda_{q,t}|^4 \mathbb{E}_U \left[ |\langle \phi_{q,t} | U | \psi \rangle|^4 \right] 
+ 2 \sum_{q,t \neq q',t'} |\lambda_{q,t}|^2 |\lambda_{q',t'}|^2  \mathbb{E}_U \left[ |\langle \phi_{q,t} | U | \psi \rangle|^2  |\langle \phi_{q',t'} | U | \psi \rangle|^2 \right] \nn
& \leq
2 \sum_{q,t,q',t'} |\lambda_{q,t}|^2 |\lambda_{q',t'}|^2  \mathbb{E}_U \left[ |\langle \phi_{q,t} | U | \psi \rangle|^2  |\langle \phi_{q',t'} | U | \psi \rangle|^2 \right] \nn
& \leq 2 \sum_{q,t,q',t'} |\lambda_{q,t}|^2 |\lambda_{q',t'}|^2  
\sqrt{ \mathbb{E}_U \left[ |\langle \phi_{q,t} | U | \psi \rangle|^4 \right] 
\mathbb{E}\left[ \langle \phi_{q',t'} | U | \psi \rangle|^4 \right] 
} \nn
& = 2 \left( \sum_{q,t} |\lambda_{q,t}|^2   
\sqrt{ \mathbb{E}_U \left[ |\langle \phi_{q,t} | U | \psi \rangle|^4 \right] 
} \right)^2 \, ,
\end{align}
where the second inequality is an application of Cauchy-Schwarz.

In conclusion we obtain the upper bound
\begin{align}
\gamma \leq 
2 \max_\phi \left( 
\frac{
\sum_{q,t} |\lambda_{q,t}|^2   
\sqrt{ \mathbb{E}_U \left[ |\langle \phi_{q,t} | U | \psi \rangle|^4 \right]}
}{ 
\sum_{q,t} |\lambda_{q,t}|^2   
\mathbb{E}_U \left[ |\langle \phi_{q,t} | U | \psi \rangle|^2 \right] 
}
\right)^2 
\leq 2 \max_{q,t}
\frac{
\mathbb{E}_U \left[ |\langle \phi_{q,t} | U | \psi \rangle|^4 \right]
}{ 
\mathbb{E}_U \left[ |\langle \phi_{q,t} | U | \psi \rangle|^2 \right]^2 
} \, .
\label{gamma1}
\end{align}
Compared with the definition in Eq.\ (\ref{gamma0}), to compute this upper bound we only need to consider the basis vectors $\phi_{q,t}$. Furthermore, by symmetry, the quantities 
$\mathbb{E}_U \left[ |\langle \phi_{q,t} | U | \psi \rangle|^4 \right]$ and
$\mathbb{E}_U \left[ |\langle \phi_{q,t} | U | \psi \rangle|^2 \right]$ are independent of $t$.
The expression in Eq.\ (\ref{gamma1}) is therefore suitable to be evaluated numerically. The penalty that we pay for this simplification is the multiplication by a factor $2$.
Some examples are shown in Table \ref{tab:small}.
More numerical estimations of $\gamma$ are shown in Appendix \ref{Sec:speed}.
\color{black}
Our numerical search suggests that the maximum in Eq.~(\ref{gamma1}) is achieved for $q = (n,0,0, \dots, 0)$.
Supported by these numerical results, we formulate the following conjecture:
\begin{conjecture}\label{con:gamma}
The maximum in Eq.\ (\ref{gamma1}) is obtained when all photons occupy the same mode, i.e., for $q = (n,0,0, \dots, 0)$. This yields
\begin{align}
\gamma 
\leq 2
\frac{
\mathbb{E}_U \left[ |\langle \phi_{q} | U | \psi \rangle|^4 \right]
}{ 
\mathbb{E}_U \left[ |\langle \phi_{q} | U | \psi \rangle|^2 \right]^2 
} \, ,
\end{align}
where
\begin{align}
| \phi_{q} \rangle = |n,0,0,\dots, 0 \rangle \, .
\end{align}
\end{conjecture}
This conjecture is used to produce Fig.\ \ref{Fig:rates}.
\color{black}

\begin{table}[t!]\centering
\begin{tabular}{ |c|  c| c| c|}
\hline
$(m,n)$  & Photon pattern &  $2\gamma_q$      &  $2(n+1)$ \\ 
\hline\hline
$(6,2)$ & $(1,1,0,0,0,0)$  &  $3.770$ &  $6$ \\
      &   $(2,0,0,0,0,0)$   &   $4.314$ &   \\ 
\hline
$(10,2)$ & $(1, 1,0,0,0,0,0,0,0,0)$  &   $4.256$ & 6\\
       &   $(2,0,0,0,0,0,0,0,0,0)$   &   $5.136$  & \\ 
\hline
$(20,2)$ & $(1,1,0,0,\dots)$   &  $4.751$ &  6\\
         & $(2,0,0,0,\dots)$    & $5.894$ &  \\
\hline

(40,2) 
& $ (1, 1,0,0,\dots) $   &  $4.593$ & 6 \\
& $ (2,0,0,0,\dots) $   &  $5.882$ & \\
\hline

$(10,3)$ & $(1, 1, 1,0,\dots)$ & $4.562$  &  $8$  \\
       & $(1, 2,0,0,\dots)$  &  $5.366$ &  \\
     & $(3,0,0,0,\dots)$    & $6.968$  & \\ 
\hline
     %
(40,3)
& $ (1, 1, 1,0,\dots) $   &  $5.475$ & 8\\
& $ (1, 2,0,0, \dots) $   &  $6.853$ &\\
& $ (3,0,0,0, \dots) $   &  $9.717$ &  \\
\hline
\end{tabular}
\caption{\label{tab:small} The table shows numerically computed values of $\gamma_q$ for different number of photons ($n$), modes ($m$), and photon occupancy patterns. The upper bound $ \gamma \leq 2(n+1)$ is shown here for comparison, but it holds only in the regime of $m \gg n^2 \gg 1$.}
\end{table}

\section{Results}\label{sec:results}

Our main result is an estimate of the minimum key size $K_\epsilon$ that guarantees that the accessible information $I_\mathrm{acc}(X;B)$ is of order $\epsilon$.
\textcolor{black}{This estimate is expressed in terms of the parameters $c_\mathrm{min}$ and $\gamma$ introduced in Section \ref{Sec:prel}.}

\begin{proposition}\label{Prop_1}

Consider the QDL protocol described in Section \ref{Sec:Protocol}, which encodes $\log{M}$ bits of information using $n$ photons over $m$ modes.
For any $\epsilon, \xi \in (0,1)$, and for any $K > K_\epsilon$, there exist choices of $K$ linear optics unitaries such that $I_\mathrm{acc}(X;B) < 2 \epsilon \log{\frac{1}{c_\mathrm{min}}}$, where
\begin{align}\label{K-1}
K_\epsilon =
\max \left\{
\gamma \left[ \frac{256}{\epsilon^3} \frac{d}{M} \ln{\left(\frac{20}{\epsilon c_\mathrm{min}} \right)} + \frac{32}{ \epsilon^2} \ln{M} \right] 
,
\frac{32}{\epsilon^2} \frac{(\ln{2d})^2}{M c_\mathrm{min}}  
\right\}
\end{align}
and $M = \xi C$.
Recall that $d = {n+m-1 \choose n}$ is the dimension of the Hilbert space with $n$ photons over $m$ modes, and $C = { m \choose n }$ is the number of states with no more than one photon per mode. 

\end{proposition}

%


The parameters $\gamma$ and $c_\text{min}$ depend on the particular values of $n$ and $m$.
We identify three regimes for $n$ and $m$:
\begin{enumerate}

\item For $n=1$, the group of linear optical passive unitaries spans all unitaries in the subspace of $n=1$ photon over $m$ modes. The single-photon representation of the group of LOP unitaries is the fundamental representation of $U(m)$.
We then obtain $\gamma = 2$ and $c_{\min} =1/m$ \cite{PRL,NJP}. 

\item In the regime where $m \gg n^2 \gg 1$ (the no-collision regime), $\gamma \approx 2(n+1)$ and $c_\text{min} \approx n!/m^n$ (see Appendix \ref{Sec:speed}.
 
\item For generic values of $n$ and $m$, to the best of our knowledge both $\gamma$ and $c_\text{min}$ need to be calculated numerically. The estimation can be simplified if we assume Conjectures \ref{con:cmin} and \ref{con:gamma} introduced in Sec.\ \ref{Sec:prel}.
\end{enumerate}

\begin{figure}[t!] \centering
 \includegraphics[trim = 0cm 0cm 0cm 0cm, clip, width=0.55\linewidth]{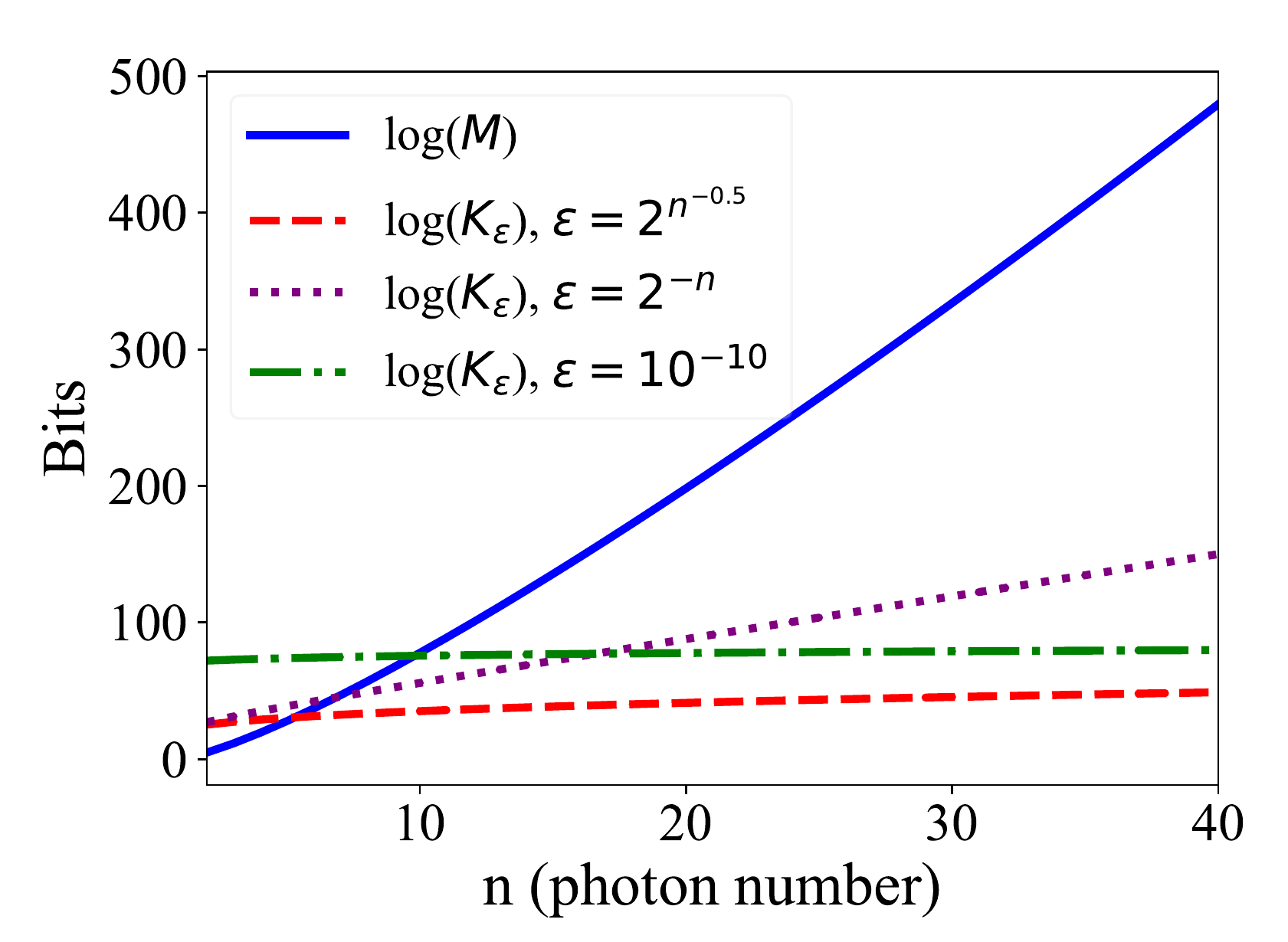} 
 \caption{ 
The solid blue line shows the number of transmitted bits $\log M =\log{\xi {m \choose n}}$ versus the photon number $n$, for a fraction of the code space $\xi = 0.01$ and $m = n^3$.
 The other lines shows the estimate of the 
 secret key consumption $\log{K}$ in Eq.~\eqref{K-1}). 
 This is obtained using $\gamma = 2(n+1)$ and $c_\mathrm{min} = n!/m^n$, i.e., assuming the values in the limit of no-collision.
The other parameters are: $\epsilon = 2^{-n^s}$, $s=0.5$ (red dashed); $s=1$ (purple dotted); 
$\epsilon=10^{-10}$ (green dotted dashed). 
If we choose the security parameter $\epsilon \propto 2^{-n^s}$, then $I_\text{acc} \rightarrow 0$ as $n\rightarrow \infty$. 
When the blue curve is higher than the other curves, the message is longer than the key. In this case, QDL beats the classical one-time pad and allows to expand the initial secret key of $\log{K_\epsilon}$ bits into a longer key of $\log{M}$ bits.
 }
 \label{f:rates}
\end{figure}

We can write Eq.~(\ref{K-1}) as 
\begin{align}\label{K-3}
\log{K_\epsilon} = \max \left\{ 
\log{\gamma} + \log{\frac{d}{M}} + f(\epsilon, c_\mathrm{min}, M) \, ,
\log{\left( \frac{1}{M c_\mathrm{min} } \right)} + g(\epsilon,c_\mathrm{min}, d)
\right\} \, ,
\end{align}
where the functions $f$ and $g$ scale as $\log{(1/\epsilon)}$.
For illustration, Fig.\ \ref{f:rates} shows $\log{M}$ and an estimate of $\log{K_\epsilon}$ as functions of $n$. To obtain the plot, we have chosen $m = n^3$ and used the limiting values for the parameters $\gamma = 2(n+1)$ and $c_\text{min} = n!/m^n$.

Note that, as $\epsilon$ is expected to be sufficiently small, 
this estimate for the secret key size is useful only in the limit of asymptotically large $K_\epsilon$, i.e., when one encodes information using asymptotically many modes and photons.
This is certainly not the regime one is willing to test in an experimental demonstration of QDL.

The QDL protocol outperforms the classical one-time pad when $\log{M} > \log{K_\epsilon}$, for some reasonably small value of $\epsilon$. Some numerical examples are in Fig.~\ref{f:rates}, which show the gap between $\log{M}$ and $\log{K_\epsilon}$ increases with increasing number of modes and photons.
For example, for $n=20$, $m=8000$, $\xi = 0.01$, and $\epsilon = 10^{-10}$, we obtain $\log{M} \simeq 192$ and $\log{K_\epsilon} \simeq 127 < 0.7 \log{M}$. This shows explicitly that we can achieve information theoretical security with a private key shorter than the message if $n$ and $m$ are large enough.

\vspace{5mm}


\noindent
\textit{Scaling up the communication protocol}:
in a practical communication scenario, not only one signal, but a large number of signals are sent from Alice to Bob through a given quantum communication channel. 
Consider a train of $\nu \gg 1$ channel uses, where Alice encodes a classical variable $X^{(\nu)}$ into tensor-product code words of the form 
\begin{align}
|\boldsymbol{\psi_x}\rangle = |\psi_{x_1}\rangle \otimes  |\psi_{x_2}\rangle \otimes \dots |\psi_{x_\nu}\rangle \, ,
\end{align}
where each component $\psi_{x_1}$ is a state of $n$ photons over $m$ modes.
Over $\nu$ channel uses, the total number of code words is denoted as $M^{(\nu)} = \xi C^n$, and the code rate is $\lim_{\nu \to \infty} \frac{1}{\nu} \log{M^{(\nu)}} = \log{C}$.
Similarly, Alice applies local unitaries to these code words,
\begin{align}
\boldsymbol{U_k} = U_{k_1} \otimes U_{k_2} \dots \otimes U_{k_\nu} \, ,
\end{align}
for a total number of $K^{(\nu)}$ allowed unitaries acting on $\nu$ channel uses.
We denote as $B^\nu$ the outputs of $\nu$ channel uses received by Bob.
The security condition on the mutual information then reads $I_\mathrm{acc}(X^\nu;B^\nu) = O(\epsilon \log{M^{(\nu)}} )$.

\begin{corollary} \label{Corol_1}

For any $\epsilon, \xi \in (0,1)$, and for any $K^{(\nu)} > K_\epsilon$, there exist choices of $K^{(\nu)}$ local unitaries upon $\nu$ channel uses such that $I_\mathrm{acc}(X^\nu;B^\nu) \leq 2 \epsilon \log{ \frac{1}{c_\mathrm{min}^n } }$, where
\begin{align} \label{K-4}
K_\epsilon = \max\left\{ 
\gamma^\nu \left[ \frac{512}{\epsilon^3} \frac{d^\nu}{M^{(\nu)}} \ln{\left(\frac{20}{\epsilon c_\mathrm{min}^\nu} \right)} 
                                     + \frac{64}{\epsilon^2} \ln{M^{(\nu)}} \right] 
,
 \frac{32}{\epsilon^2} \frac{(\ln{2d^\nu})^2}{M^{(\nu)} c_\mathrm{min}^\nu}  
\right\} \, .
\end{align}
The minimum secret key consumption rate then reads
\begin{align} \label{K-5}
k := \lim_{\nu\to\infty} \frac{1}{\nu} \log{K_\epsilon} =
\max\left\{ 
\log{\gamma} + \log{\frac{d}{C}}  
,
\log{\frac{1}{C c_\mathrm{min}}}
\right\} \, .
\end{align}

\end{corollary}

Corollary \ref{Corol_1} allows us to estimate the net secret key rate as the difference between the code rate and the secret key consumption rate, 
\begin{equation}
    r_\mathrm{QDL} = \log{C} - k 
    \, ,
\end{equation}
where conjecture \ref{con:cmin} implies $k= \log{\gamma} + \log{\frac{d}{C}}$.
If $r_\mathrm{QDL} > 0$, then the QDL is successful in beating the classical one-time pad and generates a secret key at a rate of $\log{C}$ bits per channel use larger than the key consumption rate of $k$ bits.
\color{black}

We can compare these results with the classical one-time pad encryption as well as previously known QDL protocols. 
We consider the three parameters that characterise symmetric key encryption: the length $\log{K}$ of the initial secret key, the length $\log{M}$ of the message, and the security parameter $\epsilon$.
Classical one-time pad requires $\log{K} = \log{M}$ for perfect encryption ($\epsilon =0$). Therefore, the comparison with QDL makes sense in the regime where $\epsilon$ can be made arbitrary small.
In this regime, we can then say that a QDL protocol beats the classical one-time pad if $K \ll M$.

The QDL protocol that has up to now the largest gap between $K$ and $M$ was proposed by Fawzi et al.\ in Ref.\ \cite{Fawzi13}. This protocol requires an initial key of constant size $\log{K} \sim \log{1/\epsilon}$ for any sufficiently large $M$.
This is obtained by using random unitaries in the $M$-dimensional Hilbert space, and therefore requires a universal quantum computer acting on a large Hilbert space.

Proposition \ref{Prop_1} shows that there exist QDL protocols with $\log{K} \sim O(\log{1/\epsilon}) + \log{(d/M)} = O(\log{1/\epsilon}) + \log{(d/C)} + \log{(1/\xi)}$. Comparing with Ref.\ \cite{Fawzi13}, the length of the secret key has an overhead proportional to 
\begin{align}
\log{\frac{d}{M}} = \log{\frac{(m+n+1)!(m-n)!}{m! (m+1)!}} \, .
\end{align}
The advantage with respect to Ref.\ \cite{Fawzi13} is that the encryption only requires linear optical passive unitaries.
For $m$ and $n$ large, using the Stirling approximation we obtain
\begin{align}
\log{\frac{d}{M}} & \sim n \log{\left( 1 + \frac{2n}{m-n} \right)} + m \log{\left( 1 - \frac{n^2}{m^2} \right)}  \, ,
\end{align}
which becomes negligibly small in the limit of diluted photons, $m \gg n^2 \gg 1$. 

Corollary \ref{Corol_1} shows the existence of QDL protocols for $\nu$ channel uses where a secret key of $\log{K} \sim \nu \left( \log{\gamma} + \log{d/C} \right)$ allows us to encrypt $\log{M} \sim \nu \log{C}$, where $\epsilon \to 0$ in the limit that $\nu \to \infty$, and the constant $\gamma$ depends on the particular choice of the parameters $n$ and $m$. Note that in these protocols the secret key length $\log{K}$ is not constant, but scales as the message length $\log{M}$. Although they have the same scaling, we can still have $\log{M} > \log{K}$ in some regime. Despite being less efficient in terms of key use, the advantage of these protocols is that they only need linear optics passive unitaries acting on a small number of photons and modes, i.e., $n$ and $m$ can be chosen finite and small.
%
For example, for $n=10$ photons over $m=30$ modes, we obtain $\log{M} \simeq 25$ and $\log{(d/M)} \simeq 4.4 < \frac{1}{5} \log{M}$. From table \ref{table:conj} we also obtain the numerical estimates $\log{\gamma} < \log{(111.5)} \simeq 6.8 < \frac{1}{3} \log{M}$. Putting $k = \lim_{\nu\to\infty} \frac{1}{\nu} \log{K}$, we obtain the following estimate for the asymptotic rate of secret key consumption,
\begin{align}
k = \log{\gamma} + \log{\frac{d}{M}} \simeq 
4.4 + 6.8 = 11.2 < \frac{1}{2} \log{M} \, .
\end{align}
This shows explicitly that less than $\log{M}$ bits of secret key are used to encrypt a message of $\log{M}$ bits.
Therefore, the net key generation rate in this case is 
\begin{align}
r_\mathrm{QDL} = \log{M} - k > \frac{1}{2}\log{M} \, .
\end{align}
In Section \ref{Sec:real} we consider the effect of photon loss in terms of the net rate per mode, $r_\mathrm{QDL}/m$.
\color{black}

\section{Proof of Proposition \ref{Prop_1}} 
\label{Sec:Proof1}

We prove the proposition using a random-coding argument.
We show that a random choice of the code and of the set of scrambling unitaries leads, with high probability, to a QDL protocol that satisfies the security property.

The code book $\bar{\mathcal{C}}_n^m$ of cardinality $M$ is randomly chosen by sampling from the code book $\mathcal{C}_n^m$ of cardinality $C$.
We put $M = \xi C$.
For $\xi \ll 1$, we expect that the $M$ code words are all distinct up to terms of second order in $\xi$. Therefore the $M$ code words encode $\log{M} - O(\log{(1/\xi)})$ bits of information. 

The sender Alice first prepares a state $|\psi_x\rangle$, then applies a linear optics unitary $U_k$. The unitary is chosen among a pool of $K$ elements according to a secret key of $\log{K}$ bits.
We choose the pool of unitaries by drawing $K$ unitaries i.i.d.\ according to the uniform Haar measure on the group $U_\mathrm{LO}(m)$ of linear optics unitary transformations on $m$ modes.
If the receiver does not know the secret key, the state is described by the density operator
\begin{equation}
\rho_B^x = \frac{1}{K} \sum_{k=1}^K U_k |\psi_x\rangle \langle \psi_x| U_k^\dag \, .
\end{equation}

Given the classical-quantum state
\begin{equation}
\rho_{XB} = \frac{1}{M} \sum_{x=1}^{M} | x \rangle \langle x | \otimes \rho_B^x \, ,
\end{equation}
Bob attempts to extract information from this state by applying a measurement $M_{B \to Y}$. 
Such a measurement is characterised by the POVM elements $\{ \alpha_y |\phi_y\rangle \langle \phi_y| \}_y$, where $\phi_y$'s are unit vectors and $\alpha_y >0$ such that $\sum_y \alpha_y |\phi_y\rangle \langle \phi_y| = \mathbb{I}$, with $\mathbb{I}$ the identity operator.
Without loss of generality we can consider rank-one POVM only \cite{DiVin}.
The output of this measurement is a random variable $Y$ with probability density
\begin{equation}
p_{Y}(y) = \alpha_y  \left\langle \phi_y \left| \sum_{x=1}^{M} \frac{1}{M} \rho_B^x \right| \phi_y \right\rangle  \, ,
\end{equation}
and conditional probability
\begin{equation}
p_{Y|X=x}(y) = \alpha_y \langle \phi_y | \rho_B^x | \phi_y \rangle  \, .
\end{equation}

The accessible information is the maximum mutual information between $X$ and $Y$:
\begin{equation}
I_\mathrm{acc}(X;B) = \sup_{M_{E \to Y}} I(X;Y) \, ,
\end{equation}
where
\begin{align}
I(X;Y) & =  H(X) + H(Y) - H(XY) = H(Y) - H(Y|X) \nn
 &= - \sum_y p_Y(y) \log{p_Y(y)}
    + \sum_{x=1}^{M} \frac{1}{M} \sum_y p_{Y|X=x}(y) \log{[p_{Y|X=x}(y)]} \nn
 &= - \sum_y \alpha_y \left \langle \phi_y \left| \sum_{x=1}^{M} \frac{1}{M} \rho_B^x \right| \phi_y \right\rangle \log{\left(\alpha_y \langle \phi_y | \sum_{x=1}^{M} \frac{1}{M} \rho_B^x | \phi_y \rangle \right)} \nn
   & \quad + \sum_{x=1}^{M} \frac{1}{M} \sum_y \alpha_y \langle \phi_y | \rho_B^x | \phi_y \rangle \log{\left[\alpha_y \langle \phi_y | \rho_B^x | \phi_y \rangle \right]} \, .
\end{align}
This yields 
\begin{align}\label{Iacc1shot}
I(X;Y) =
\sum_y \alpha_y & \left\{ 
 - \left \langle \phi_y \left| \frac{1}{M} \sum_{x=1}^{M} \rho_B^x \right| \phi_y \right \rangle 
\log{ \left \langle \phi_y \left| \frac{1}{M} \sum_{x=1}^{M} \rho_B^x \right| \phi_y \right \rangle } \right. \nn
& + \left. \frac{1}{M} \sum_{x=1}^{M} \langle \phi_y| \rho_B^x | \phi_y \rangle \log{ \langle \phi_y| \rho_B^x | \phi_y \rangle }
\right\} \, .
\end{align}

Note that the accessible information is written as the difference of two entropy-like quantities. 
The rationale of the proof is to show that for $K$ large enough, and for random choices of the unitaries and of the code words, both terms in the curly brackets are arbitrarily close to 
\begin{align}
- \langle \phi_y | \bar{\rho}_B | \phi_y \rangle \log{ \langle \phi_y | \bar{\rho}_B | \phi_y \rangle } 
\end{align}
for all vectors $\phi_y$, where $\bar\rho_B$ is as in Eq.\ (\ref{barrhoE}). 
This in turn implies that the accessible information can be made arbitrarily small. 
To show this we exploit the phenomenon of concentration towards the average of the sum of i.i.d.\ random variables. This concentration is quantified by concentration inequalities.


We now proceed along two parallel directions.


First, we apply the matrix Chernoff bound \cite{Chernoff} to show that $\frac{1}{M} \sum_{x=1}^{M} \rho_B^x$ is close to $\bar\rho_B$.
In particular the matrix Chernoff bound implies that the inequality
\begin{equation}
\frac{1}{M} \sum_{x=1}^{M} \rho_B^x \leq (1+\epsilon) \bar{\rho}_B
\end{equation}
holds true up to a failure probability
\begin{align}\label{pfail1_main}
p_1 \leq \exp{\left[ \ln{(2d)} - \frac{\epsilon}{4} \sqrt{\frac{M K c_\mathrm{min}}{2}} \right] } \, .
\end{align}
This in turn implies
\begin{align}
- \left \langle \phi \left| \frac{1}{M} \sum_{x=1}^{M} \rho_B^x  \right | \phi \right \rangle 
\log{ \left \langle \phi \left| \frac{1}{M} \sum_{x=1}^{M} \rho_B^x  \right| \phi \right\rangle }
& \leq - (1+\epsilon)\langle \phi | \bar{\rho}_B | \phi \rangle \log{ \langle \phi | \bar{\rho}_B | \phi \rangle } 
\label{Conc1}
\end{align}
uniformly for all $\phi$.
The details are presented in Appendix \ref{Sec:Chernoff} below.

Second, we apply a tail bound from A. Maurer \cite{AM} to show that
\begin{align}
\langle \phi | \rho_B^x | \phi \rangle \geq (1-\epsilon)
\langle \phi | \bar{\rho}_B | \phi \rangle \, ,
\end{align}
up to a failure probability
\begin{align}\label{pfail2_main}
p_2 \leq \exp{\left[
2d \ln{\left( \frac{20}{\epsilon c_\mathrm{min}} \right)}
+ \frac{\epsilon M}{4} \ln{M}
- \frac{ KM \epsilon^3 }{128\gamma} \right] }  \, .
\end{align}
The above applies uniformly to all unit vectors $\phi$ and for almost all values of $x$. This implies that
\begin{align}
\langle \phi| \rho_B^x | \phi \rangle \log{ \langle \phi| \rho_B^x | \phi \rangle }
\leq (1-\epsilon )\langle \phi| \bar{\rho}_B | \phi \rangle \log{ (1-\epsilon)\langle \phi| \bar{\rho}_B | \phi \rangle } \, .
\end{align}
In conclusion, we obtain that, up to a probability smaller than $p_2$,
\begin{align}
\frac{1}{M} \sum_{x=1}^{M} \langle \phi| \rho_B^x | \phi \rangle \log{ \langle \phi| \rho_B^x | \phi \rangle }
& \leq (1-\epsilon )\langle \phi| \bar{\rho}_B | \phi \rangle \log{ \langle \phi| \bar{\rho}_B | \phi \rangle }
\, . 
\label{Conc2}
\end{align}
The details are presented in Appendix \ref{Sec:Maurer}.



Putting the above results in Eq.\ (\ref{Conc1}) and (\ref{Conc2}) into Eq.\ (\ref{Iacc1shot}) 
we finally obtain
\begin{align}
I(X;Y) \leq - 2 \epsilon \sum_y \alpha_y  
\langle \phi_y| \bar{\rho}_B | \phi_y \rangle \log{ \langle \phi_y| \bar{\rho}_B | \phi_y \rangle }
\, .
\end{align}
Recall that $p_Y(y) = \alpha_y \langle \phi_y| \bar{\rho}_B | \phi_y \rangle$
is a probability distribution. Therefore, as the average is always smaller that the maximum,
we obtain
\begin{align}
I(X;Y) \leq - 2 \epsilon \min_\phi \log{ \langle \phi| \bar{\rho}_B | \phi \rangle } 
= 2 \epsilon \log{ \frac{1}{c_\mathrm{min}} } \, ,
\end{align}
where $c_\mathrm{min} := \min_\phi \langle \phi| \bar{\rho}_B | \phi\rangle$ can be computed as shown in Section \ref{Sec:prel}.

The above bound on the accessible information is not deterministic, but the probability $p_1 + p_2$ that it fails can be made arbitrary small provided $K$ is large enough (see Appendix \ref{Sec:bad} for details).
This probability is bounded away from $1$ if
\begin{align}
K > \frac{32}{\epsilon^2} \frac{(\ln{2d})^2}{M c_\mathrm{min}}   \, , 
\end{align}
and 
\begin{align}
K > 128 \gamma \left[ \frac{2}{\epsilon^3} \frac{d}{M} \ln{\left(\frac{20}{\epsilon c_\mathrm{min}} \right)} 
                                     + \frac{1}{4 \epsilon^2} \ln{M} \right] \, .
\end{align}

The size of $K$ critically depends on the factor $\gamma$, which determines the convergence rate of the Maurer tail bound. How to estimate this coefficient is the subject of Appendix \ref{Sec:speed}.

\section{Proof of Corollary \ref{Corol_1}} \label{Sec:col1}



Consider a train of $\nu \gg 1$ channel uses.
Alice encodes information using $M^{(\nu)}$ code words of the form $|\boldsymbol{\psi_x}\rangle = |\psi_{x_1}\rangle \otimes  |\psi_{x_2}\rangle \otimes \dots |\psi_{x_\nu}\rangle$, where each component $\psi_{x_1}$ is chosen randomly and independently from the code book $\mathcal{C}_n^m$, which has cardinality $C$.
Each $\nu$-fold code word is uniquely identified by the multi-index $\boldsymbol{x} = x_1,x_2, \dots, x_\nu$. We put $M^{(\nu)} = \xi C^\nu$, where $\xi \ll 1$ is a small positive constant.

First Alice encodes information across $\nu$ signal uses using the code words $\boldsymbol{\psi_x}$, then she applies local unitaries $\boldsymbol{U_k} = U_{k_1} \otimes U_{k_2} \dots \otimes U_{k_\nu}$ to scramble them.
The set of possible unitaries is made of $K^{(\nu)}$ elements. 
These unitaries are chosen by sampling identically and independently from the Haar measure on the unitary group $U_\mathrm{LO}(m)$ of linear optical passive
unitary transformations on $m$ modes. 
Note that, whereas $\nu$ is arbitrary large, the number of modes $m$ in each signal transmission will be kept constant and relatively small. Also, the number of photons per channel use is fixed and equal to $n$.

The accessible information is then formally equivalent to the one in Eq.\ (\ref{Iacc1shot}):
\begin{align}
I(X^\nu;Y) = 
\sum_y \alpha_y & \left\{ 
-\left \langle \phi_y \left| \frac{1}{M^{(\nu)}} \sum_{\boldsymbol{x}=1}^{M^{(\nu)}} \rho_B^{\boldsymbol{x}} \right| \phi_y \right \rangle 
\log{ \left \langle \phi_y\left | \frac{1}{M^{(\nu)}} \sum_{\mathbf{x}=1}^{M^{(\nu)}} \rho_B^{\boldsymbol{x}} \right | \phi_y \right \rangle } \right. \nn
& + \left. \frac{1}{M^{(\nu)}} \sum_{\boldsymbol{x}=1}^{M^{(\nu)}} \langle \phi_y| \rho_B^{\boldsymbol{x}} | \phi_y \rangle \log{ \langle \phi_y| \rho_B^{\boldsymbol{x}} | \phi_y \rangle }
\right\} \, ,
\end{align}
where for each $\boldsymbol{x} = x_1, x_2, \dots, x_\nu$,
\begin{align}
\rho_B^{\boldsymbol{x}} & = \sum_{\boldsymbol{k}=1}^{K^{(\nu)}} \boldsymbol{U_k} |\boldsymbol{\psi_x} \rangle\langle \boldsymbol{\psi_x}| \boldsymbol{U_k}^\dag \\
                  & = \sum_{k_1,\dots,k_\nu=1}^{K^{(\nu)}} \left( U_{k_1} \otimes \dots \otimes U_{k_\nu} \right) 
                  \left( |\psi_{x_1}\rangle\langle \psi_{x_1}| \otimes \dots \otimes |\psi_{x_\nu}\rangle\langle \psi_{x_\nu}| \right)
                  \left( U_{k_1}^\dag \otimes \dots \otimes U_{k_\nu}^\dag \right) \\
                  & = \bigotimes_{i=1}^\nu \sum_{k_i=1}^{K^{(\nu)}} U_{k_i} |\psi_{x_i}\rangle\langle \psi_{x_i}| U_{k_i}^\dag \, .
\end{align}

This in particular implies
\begin{align}
\bar{\rho}_B^{(\nu)} := \mathbb{E}_{\boldsymbol{U}}[\boldsymbol{U_k} |\boldsymbol{\psi_x}\rangle\langle\boldsymbol{\psi_x}| \boldsymbol{U_k}^\dag] 
= \bar{\rho}_B^{\otimes \nu} \, ,
\end{align}
and therefore
\begin{align}
c^{(\nu)}_\mathrm{min} & := \min_\phi \langle \phi| \bar{\rho}_B^{(\nu)} | \phi \rangle 
= \min_\phi \langle \phi| \bar{\rho}_B^{\otimes\nu} | \phi \rangle
= c^\nu_\mathrm{min}\, \\ 
\gamma^{(\nu)} 
& : = \max_\phi \frac{ \mathbb{E}_{\boldsymbol{U}}[|\langle \phi | \boldsymbol{U_k} | \boldsymbol{\psi_x} \rangle |^4] }{ \mathbb{E}_{\boldsymbol{U}}[|\langle \phi | \boldsymbol{U_k} | \boldsymbol{\psi_x} \rangle |^2]^2 } \, .
\end{align}

To bound $\gamma^{(\nu)}$ we can first decompose a generic vector $\phi$ as $|\phi\rangle = \sum_{\boldsymbol{q},\boldsymbol{t}} \alpha_{\boldsymbol{q},\boldsymbol{t}} |\phi_{\boldsymbol{q},\boldsymbol{t}}\rangle$, 
where $|\phi_{\boldsymbol{q},\boldsymbol{t}} \rangle = |\phi_{q_1,t_1}\rangle \otimes |\phi_{q_2,t_2}\rangle \otimes \cdots |\phi_{q_\nu,t_\nu}\rangle$ define a basis of product vectors. 
We can then apply the Cauchy-Schwarz inequality, as shown in Section \ref{Sec:prel}, to obtain the upper bound $\gamma^{(\nu)} \leq 2 \gamma^\nu$.

In conclusion, we can straightforwardly repeat the proof of Section \ref{Sec:Proof1} with these new parameters. This yields that, for any arbitrarily small $\epsilon$, the bound
\begin{align}
I(X^\nu;Y) \leq 2 \epsilon \log{ \frac{1}{c_\mathrm{min}^{\nu}} } \, ,
\end{align}
holds with non-zero probability provided that (recall that $M^{(\nu)} = \xi C^\nu$)
\begin{align}
K > \max\left\{ 
\gamma^\nu \left[ \frac{512}{\epsilon^3} \frac{d^\nu}{M^{(\nu)}} \ln{\left(\frac{20}{\epsilon c_\mathrm{min}^\nu} \right)} 
                                     + \frac{64}{\epsilon^2} \ln{M^{(\nu)}} \right] 
,
 \frac{32}{\epsilon^2} \frac{(\ln{2d^\nu})^2}{M^{(\nu)} c_\mathrm{min}^\nu}  
\right\} \, .
\end{align}

Finally, in the limit of $\nu \gg 1$, and since $\lim_{\nu\to\infty} \frac{1}{\nu} \log{M^{(\nu)}} = \log{C}$, we obtain
\begin{align}
\lim_{\nu\to\infty} \frac{1}{\nu} \log{K_\epsilon} \geq 
\max\left\{
\log{\gamma} + \log{\frac{d}{C}}
,
\log{\frac{1}{C c_\mathrm{min}}}
\right\} \, .
\end{align}

\section{Noisy channels}\label{Sec:real}

\begin{figure}[t]\centering
\includegraphics[trim = 0cm 0cm 0cm 0cm, clip, width=0.5\linewidth]{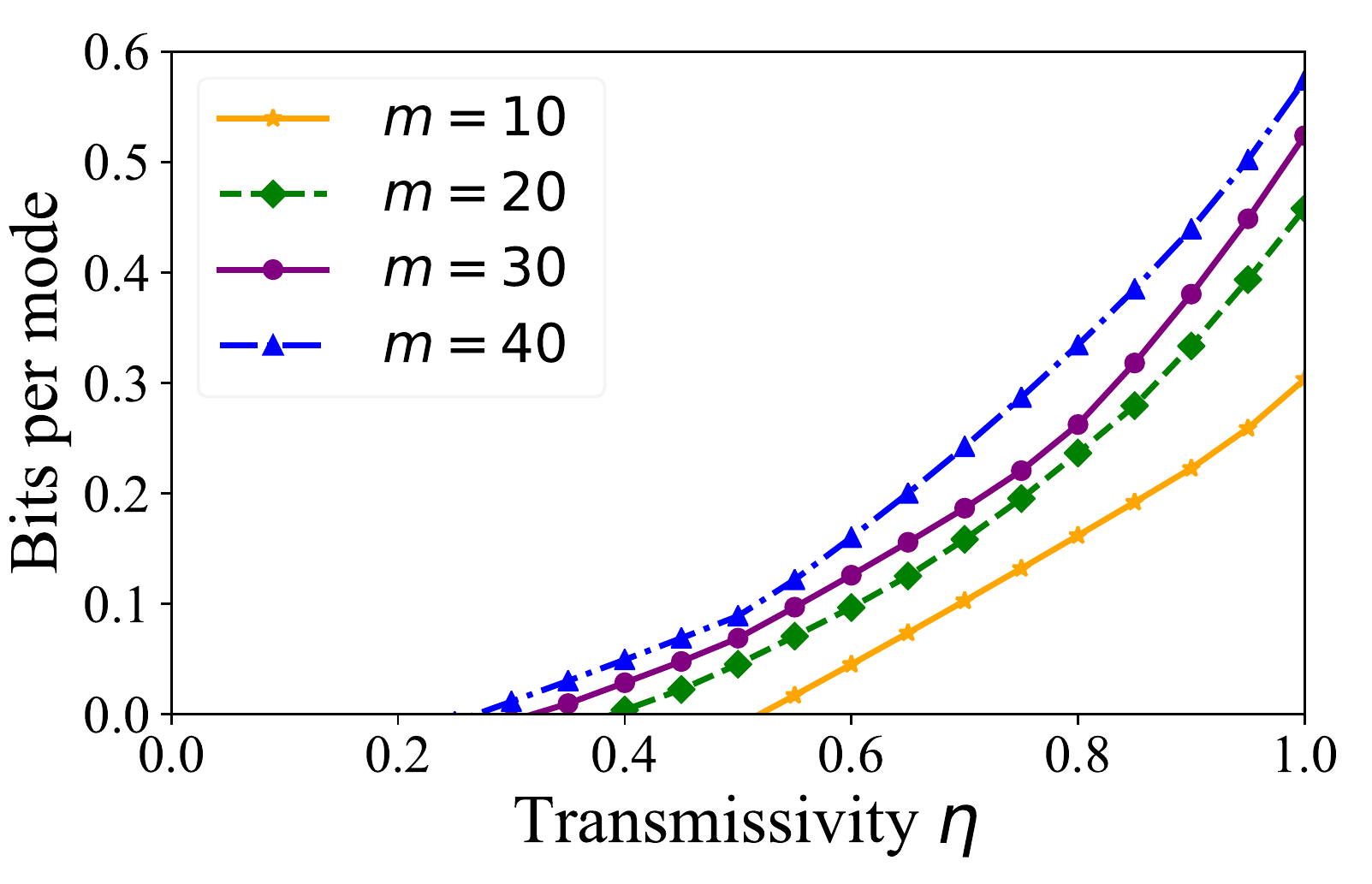} 
  \caption{
The rate-loss trade-off for our protocol with $m = 10,20,30$, and $40$. 
 The rates represent the excess number of transmitted (secure) bits per optical mode ($\frac{1}{m}\times$ Eq.~(56)) over the classical one-time pad, in the presence of loss. 
 A positive rate expresses the fact that the QDL protocol allows us to generate more secret bits than it consumes, hence beating the classical one-time pad encryption.
  The estimates of the parameters $\gamma$ and $c_\mathrm{min}$ are obtained by assuming 
  Conjectures \ref{con:cmin} and \ref{con:gamma}.
   We see that the information density per mode increases as $m$ increases.
We have  chosen $n$ to maximise the rate. The optimal value of $n$ depends on $\eta$, and $n \approx m/3$ for $\eta \approx 1$. 
  For moderate losses, the optimal $n$ decreases. 
This suggests that QDL may be observed with high loss by increasing the number of modes.
  These values for the number of photons and modes are similar to those of a recent experimental demonstration of Boson Sampling \cite{JWP102019}.}
  \label{Fig:rates}
\end{figure}

A practical communication protocol needs to account for loss and noise in the communication channel. This requires us to introduce error correction in the classical post-processing. We address this issue here and show that the structure of our proof encompasses a large class of error correcting protocols.


In the case of a noisy and lossy channel, Alice and Bob can still use the channel by employing error correction. Error correction comes with an overhead that reduces the maximum communication rate from $\log{M}$ (the maximum amount of information that can be conveyed through a noiseless channel) to $I(X;Y|K) \leq  \log{M}$, where $I(X;Y|K)$ is the mutual information given that both Alice and Bob know the secret key $K$.

\color{black}

The amount of loss and noise in the communication channel can be experimentally determined with the standard tools of {\it parameter estimation}, a routine commonly employed in quantum key distribution. This in turn allows Alice and Bob
to quantify $I(X;Y|K)$. 

In principle, error correction allows Alice and Bob to achieve a communication rate arbitrarily close to $I(X;Y|K)$. In practice, however, we can only partially achieve this goal. To model this fact, one usually introduce the error correction efficiency factor $\beta \in (0,1)$. Putting this together with Corollary \ref{Corol_1}, we obtain our estimate for the net rate of the protocol:
\begin{align}\label{eq:r}
r_\mathrm{QDL} = \beta I(X;Y|K)
- \max\left\{ 
\log{\gamma} + \log{\frac{d}{C}}  
,
\log{\frac{1}{C c_\mathrm{min}}}
\right\} \, ,
\end{align}
where a positive net rate expresses the fact that the QDL protocol allows us to expand the initial secret key into a longer one.

As an example, consider the case where Alice and Bob communicate through a lossy optical channel. The efficiency factor $\eta \in (0,1)$ represents the probability that a photon sent by Alice is detected by Bob, including both channel losses and detector efficiency. The mutual information 
$I(X;Y|K)$ between Alice and Bob can be computed explicitly (see Appendix \ref{Sec:loss} for detail). 
We obtain :
\begin{align}
I(X;Y|K) & = - \sum_{k=0}^n {n \choose k} \eta^{k} (1-\eta)^{n-k} \log{\frac{{ m-k\choose n-k}}{{ m\choose n}}} \\
& = \log{{ m\choose n}} - \sum_{k=0}^n {n \choose k} \eta^{k} (1-\eta)^{n-k} \log{{ m-k\choose n-k}} \, .
\label{eq:iabloss}
\end{align}

%

Fig.~\ref{Fig:rates} shows the quantity $r_\mathrm{QDL}/m$, i.e., the number of bits \textit{per mode}, for $\beta = 1$, for a pure loss channel with transmissivity $\eta$. 
The plot is obtained assuming Conjectures \ref{con:cmin} and \ref{con:gamma}.
This shows that QDL can be demonstrated experimentally with loss and inefficient detectors. In particular, higher loss can be tolerated by increasing the number of optical modes.
Note that the values for the number of photons and modes used to obtain this figure have been achieved experimentally in Ref.\ \cite{JWP102019}.

%
\color{black}

\section{Conclusions} \label{Sec:conclusion}

The phenomenon of Quantum Data Locking (QDL) represents one of the most remarkable separations between classical and quantum information theory.
In classical information theory, information-theoretic encryption of a string of $N$ bits can be only made by exploiting a secret key of at least $N$ bits. This is realised, for example, by using a one-time pad.
By contrast, QDL shows that, if information is encoded into a quantum system of matter or light, it is possible to encrypt $N$ bits of information with a secret key of $k \ll N$ bits.
QDL is a manifestation of the uncertainty principle in quantum information theory \cite{Wehner_2010,Coles}.

Initial works on QDL have focused on abstract protocols defined in a Hilbert space of asymptotically large dimensions. 
More recent works have extended QDL to system of relatively small dimensions that are transmitted through many uses of a communication channel. This approach allowed to incorporate error correction and led to one of the first experimental demonstrations of QDL in an optical setup \cite{Lum}.

Inspired by Boson Sampling \cite{aaronson2011computational,aaronson2014bosonsampling}, in this work we have further extended QDL to a setup where information is encoded using multiple photons scattered across many modes, and processed using linear passive optics. 
The extension of QDL to multiphoton states is technically challenging due to role played by higher-order representations of the unitary group.

Our protocols for multiphoton QDL 
has the potential to data-lock more bits per optical mode, hence can achieve a higher information density. 
Experimental realisations of our protocols are challenging but feasible with state-of-the-art technology. This is suggested by recent results in photon generation and advances in integrated linear optics, e.g., Ref.\ \cite{JWP102019} reported interference of $20$ photons across $60$ modes.

Several works have attempted to apply the physical insights of Boson Sampling in a quantum information framework beyond its defining problem. 
In this paper, we provide a protocol for quantum cryptography based on the physics of Boson Sampling.
We have presented an information-theoretic proof that a linear-optical interferometer, fed with multiple photons, is useful for quantum cryptography. 
The security of our protocol does not rely on the classical computational complexity of Boson Sampling. 
Therefore it holds for any number of modes $m$ and photon number $n$.
The security proof is based on QDL and random coding techniques. 
We have shown that our protocol remains secure when we use classical error correction to protect the channel against photon loss and other errors. It is therefore a scalable and efficient protocol for quantum cryptography.

\begin{acknowledgments}
ZH would like to thank Professor Jonathan P. Dowling for his encouragement and enthusiastic support throughout the years.
PK would like to thank Jon Dowling for being a great mentor.
ZH thanks Ryan L. Mann for insightful discussions. 
DWB is funded by Australian Research Council Discovery Projects DP160102426 and DP190102633.
JPD received support from the Air Force Office of Scientific Research, the Army Research Office, the Defense Advanced Research Projects Agency, and the National Science Foundation. 
JPD was grateful to LU Chaoyang for interesting and useful discussions.
ZH, CP, and PK are funded by the EPSRC Quantum Communications Hub, Grant No. EP/M013472/1. 
\end{acknowledgments}


\onecolumn\newpage
\appendix

\appendix

\widetext
\section{Matrix Chernoff bounds}\label{Sec:Chernoff}

The matrix Chernoff bound states the following (this formulation can be obtained directly from Theorem 19 of Ref.\ \cite{Chernoff}):
\begin{theorem}\label{Chernoff}
Let $\{ X_t \}_{t=1,\dots,T}$ be $T$ i.i.d.\ $d$-dimensional Hermitian-matrix-valued random variables, with $X_t \sim X$, 
$0 \leq X \leq R$, and $c_\mathrm{min} \leq \mathbb{E}[X] \leq c_\mathrm{max}$.
Then, for $\delta \geq 0$:
\begin{align}
\mathrm{Pr} \left\{ \frac{1}{T} \sum_{t=1}^T X_t 
\not\leq (1+\delta) \mathbb{E}[X] \right\} 
\leq d \exp{
\left\{
- T D\left[ 
(1+\delta)\frac{c_\mathrm{min}}{R} \left\| \frac{c_\mathrm{min}}{R} \right.
\right]
\right\}
} \, , 
\end{align}
where $\text{Pr}\{ x \}$ denotes the probability that the proposition $x$ is true, and
$D[u\|v] = u \ln{(u/v)} - (1-u) \ln{[(1-u)/(1-v)]}$.
\end{theorem}
Note that for $\delta>1$ we have
\begin{align}
    D\left[ 
(1+\delta)\frac{c_\mathrm{min}}{R} \left\| \frac{c_\mathrm{min}}{R} \right.
\right] \geq \frac{\delta}{4} \frac{c_\mathrm{min}}{R}\, ,
\end{align}
and for $\delta<1$
\begin{align}
    D\left[ 
(1+\delta)\frac{c_\mathrm{min}}{R} \left\| \frac{c_\mathrm{min}}{R} \right.
\right] \geq \frac{\delta^2}{4} \frac{c_\mathrm{min}}{R} \, .
\end{align}

First consider the collection of $M$ code words $\psi_x$.
We apply the Chernoff bound to the $M$ independent random variables $X_x = |\psi_x\rangle \langle \psi_x|$. 
Note that these operators are defined in a $C$-dimensional Hilbert space. For $\tau > 1$ we then have 
\begin{align}
\mathrm{Pr} \left\{ \frac{1}{M} \sum_{x=1}^{M} |\psi_x\rangle \langle \psi_x| 
\not\leq \frac{1+\tau}{C} \right\} 
\leq C \exp{
\left(
- \frac{M \tau}{4C}
\right)
} \, .
\end{align}

Consider now the collection of $K$ random variables $X_k = \frac{1}{M} \sum_x U_k |\psi_x\rangle \langle \psi_x| U_k^\dag$.
We assume that they are bounded by $R=\frac{1+\tau}{C}$.
We apply again the Chernoff bound:
\begin{align}
\mathrm{Pr} \left\{ \frac{1}{K} \sum_{k=1}^K \frac{1}{M} \sum_x U_k |\psi_x\rangle \langle \psi_x| U_k^\dag
\not\leq (1+\epsilon) \mathbb{E}[X] \right\} 
\leq  d \exp{
\left(
- \frac{C K \epsilon^2 c_\mathrm{min}}{4(1+\tau)}
\right)
}  \, . 
\end{align}
Thus the total probability reads
\begin{align}
p_1 & \leq C \exp{
\left(
- \frac{M \tau}{4C}
\right)
} + d \exp{
\left(
- \frac{C K \epsilon^2 c_\mathrm{min}}{4(1+\tau)}
\right)
} \\
& \leq C \exp{
\left(
- \frac{M \tau}{4C}
\right)
} + d \exp{
\left(
- \frac{C K \epsilon^2 c_\mathrm{min}}{8 \tau}
\right)
}
\, .
\end{align}
Putting $\tau = C \epsilon \sqrt{\frac{K c_\mathrm{min}}{2 M}}$ we obtain
\begin{align}\label{pfail1}
p_1 \leq (C+d) \exp{\left( - \frac{\epsilon}{4} \sqrt{\frac{M K c_\mathrm{min}}{2}}\right) } 
\leq 2d \exp{\left( - \frac{\epsilon}{4} \sqrt{\frac{M K c_\mathrm{min}}{2}}\right) } \, .
\end{align}
In conclusion we have obtained that, up to a probability smaller than $p_1$,
\begin{align}
\frac{1}{KM} \sum_{k=1}^K \sum_{x=1}^{M} U_k | x \rangle \langle x | U_k^\dag 
= \frac{1}{M} \sum_{x=1}^{M} \rho_B^x 
\leq (1+\epsilon) \bar{\rho}_B \, . 
\end{align}


\section{The Maurer tail bound}\label{Sec:Maurer}

We also need to apply the following concentration inequality due to A. Maurer~\cite{AM}:
\begin{theorem}\label{Maurer}
Let $\{ X_k \}_{k=1,\dots,K}$ be $K$ i.i.d.\ non-negative real-valued random variables, with $X_k \sim X$ and 
finite first and second moments, $\mathbb{E}[X],\mathbb{E}[X^2] < \infty$.
Then, for any $\tau > 0$ we have that
\begin{align}
\mathrm{Pr}\left\{ \frac{1}{K}\sum_{k=1}^K X_k < (1-\tau) \mathbb{E}[X]  \right\} 
\leq \exp{\left( - \frac{K\tau^2\mathbb{E}[X]^2}{2\mathbb{E}[X^2]} \right)} \, .
\end{align}
\end{theorem}

For any given $x$ and $\phi$, we apply this bound to the random variables
\begin{align}
X_k \equiv | \langle \phi | U_k | \psi_x \rangle|^2 \, .
\end{align}
Note that (see Section \ref{Sec:prel})
\begin{align}
\frac{1}{K} \sum_{k=1}^K X_k = \langle \phi | \rho^x_B | \phi \rangle \, ,
\end{align}
and 
\begin{align}
\mathbb{E}[X] = \langle \phi | \bar{\rho}_B | \phi \rangle \, .
\end{align}

The application of the Maurer tail bound then yields
\begin{align}\label{pbound1}
\mathrm{Pr}\left\{ \langle \phi | \rho^x_B | \phi \rangle < (1-\tau) \langle \phi | \bar{\rho}_B | \phi \rangle  \right\} 
\leq \exp{\left( - \frac{K\tau^2 }{2\gamma} \right)} \, ,
\end{align}
where
\begin{align}
\gamma = \max_\phi \frac{\mathbb{E}_U[X^2]}{\mathbb{E}_U[X]^2} 
       = \max_\phi \frac{\mathbb{E}_U[|\langle \phi | U | \psi_x \rangle |^4]}{\langle \phi | \bar\rho_B | \phi \rangle^2 } \, .
\end{align}
Note that, by symmetry, $\gamma$ is independent of $\psi_x$.
The calculation of $\gamma$ is presented in Appendix \ref{Sec:speed}.


\subsection{Extending to almost all code words}

The probability bound in Eq.\ (\ref{pbound1}) is about one given value of $x$. 
Here we extend it to $\ell$ distinct values $x_1, x_2, \dots, x_\ell$:
\begin{align}
\mathrm{Pr}\left\{  \forall x = x_1, x_2, \dots x_\ell,  \, \, \langle \phi | \rho^{x}_B | \phi \rangle < (1-\tau) \langle \phi | \bar{\rho}_B | \phi \rangle  \right\} 
\leq \exp{\left( - \frac{ \ell K\tau^2 }{2\gamma} \right)} \, ,
\end{align}
where we have used the fact that for different values of $x$ the variables are statistically 
independent (recall that the code words are chosen randomly and independently). 
Second, we extend to all possible choices of $\ell$ code words. This amount to a total
of ${M \choose \ell}$ events. Applying the union bound we obtain
\begin{align}\label{pbound2}
\mathrm{Pr}\left\{  \exists x_1, x_2, \dots x_\ell,  \, \, | \, \, \forall x = x_1, x_2, \dots x_\ell,  
\, \,  \langle \phi | \rho^{x}_B | \phi \rangle < (1-\tau) \langle \phi | \bar{\rho}_B | \phi \rangle  \right\} 
\leq {M \choose \ell} \exp{\left( - \frac{ \ell K\tau^2 }{2\gamma} \right)} \, .
\end{align}

This implies that, up to a probability smaller than ${M \choose \ell} \exp{\left( - \frac{ \ell K\tau^2 }{2\gamma} \right)}$,
$\langle \phi | \rho^{x}_B | \phi \rangle \geq (1-\tau) \langle \phi | \bar{\rho}_B | \phi \rangle$
for at least $M-\ell$ values of $x$. This in turn yields 
\begin{align}
\frac{1}{M} \sum_{x=1}^{M} \langle \phi | \rho^x_B | \phi \rangle \log{\langle \phi | \rho^x_B | \phi \rangle}
& \leq \frac{M-\ell}{M} (1-\tau)\langle \phi | \bar\rho_B | \phi \rangle \log{[(1-\tau)\langle \phi | \bar\rho_B | \phi \rangle]} \, .
\end{align}
Putting $\ell = \tau M$:
\begin{align}
\frac{1}{M} \sum_{x=1}^{M} \langle \phi | \rho^x_B | \phi \rangle \log{\langle \phi | \rho^x_B | \phi \rangle}
& \leq (1-\tau)^2 \langle \phi | \bar\rho_B | \phi \rangle \log{[(1-\tau)\langle \phi | \bar\rho_B | \phi \rangle]} \\
& \leq (1-\tau)^2 \langle \phi | \bar\rho_B | \phi \rangle \log{\langle \phi | \bar\rho_B | \phi \rangle} 
\\
& = (1-2\tau) \langle \phi | \bar\rho_B | \phi \rangle \log{\langle \phi | \bar\rho_B | \phi \rangle} 
+ O(\tau^2) \, .
\end{align}


\subsection{Extending to all vectors $\phi$}

The final step is to extend the result to all unit vectors.
This can be done by exploiting the notion of $\delta$-net. 
A $\delta$-net is a discrete and finite
set of vectors $\{ \phi_i \}$ on the unit sphere such that for any unit vector $\phi$ there
exists an element in the $\delta$-net such that
\begin{align}
\| \phi - \phi_i \|_1 \leq \delta \, .
\end{align}
It is known that there exist $\delta$-nets with no more than $(5/\delta)^{2d}$ elements \cite{CMP},
where $d$ is the Hilbert space dimension.
We put $\delta = \tau c_\mathrm{min}$, therefore the size of the net is $(5/\tau/c_\mathrm{min})^{2d}$.
Applying the union bound on Eq.\ (\ref{pbound2}) we then obtain
\begin{align}\label{pbound3}
& \mathrm{Pr}\left\{ \forall \phi, \exists x_1, x_2, \dots x_\ell,  \, \, | \, \, \forall x = x_1, x_2, \dots x_\ell,  
\, \,  \langle \phi | \rho^{x}_B | \phi \rangle < (1-2\tau) \langle \phi | \bar{\rho}_B | \phi \rangle  \right\} \nn
& \leq \left( \frac{5}{\tau c_\mathrm{min}} \right)^{2d} {M \choose \ell} \exp{\left( - \frac{ \ell K\tau^2 }{2\gamma} \right)} \, .
\end{align}

To conclude, we put $\epsilon = 4 \tau$ and obtain that, uniformly in $\phi$,
\begin{align}
\frac{1}{M} \sum_{x=1}^{M} \langle \phi | \rho^x_B | \phi \rangle \log{\langle \phi | \rho^x_B | \phi \rangle}
& \leq (1-\epsilon) \langle \phi | \bar\rho_B | \phi \rangle \log{\langle \phi | \bar\rho_B | \phi \rangle} 
+ O(\epsilon^2) \, .
\end{align}
The probability that this bound is violated is smaller than (recall that $\ell = \tau M = \epsilon M/4$)
\begin{align}
p_2 & =\left( \frac{5}{\tau c_\mathrm{min}} \right)^{2d} {M \choose \ell} \exp{\left( - \frac{ \ell K\tau^2 }{2\gamma} \right)} \\
& = \left( \frac{20}{\epsilon c_\mathrm{min}} \right)^{2d} {M \choose \epsilon M/4} \exp{\left( - \frac{ KM \epsilon^3 }{128\gamma} \right)}  \\
& \leq \left( \frac{20}{\epsilon c_\mathrm{min}} \right)^{2d} {M}^{\epsilon M/4} \exp{\left( - \frac{ KM \epsilon^3 }{128\gamma} \right)}  \, .
\label{pfail2}
\end{align}


\section{Probability of failure}\label{Sec:bad}

The above concentration inequalities allow us to prove that the protocol is
secure up a to certain probability. 
The {\it bad event} that the protocol is not secure occurs when either Eq.\ (\ref{Conc1}) or (\ref{Conc2}) is violated. 
The probability of the bad event is then smaller the sum of the corresponding
probabilities, which are given in Eq.\ (\ref{pfail1}) and (\ref{pfail2}) respectively. 
We therefore have
\begin{align}
P_\mathrm{fail} 
& \leq p_1 + p_2 
\leq 
2d \exp{\left( - \frac{\epsilon}{4} \sqrt{\frac{M K c_\mathrm{min}}{2}}\right) }
+ \left( \frac{20}{\epsilon c_\mathrm{min}} \right)^{2d} {M}^{\epsilon M/4} \exp{\left( - \frac{ KM \epsilon^3 }{128\gamma} \right)} \\
& = 
\exp{\left( \log{2d} - \frac{\epsilon}{4} \sqrt{\frac{M K c_\mathrm{min}}{2}} \right)}
+ \exp{\left[ 2d \log{\left(\frac{20}{\epsilon c_\mathrm{min}} \right)} + \frac{\epsilon M}{4} \log{M} - \frac{ KM \epsilon^3 }{128\gamma} \right]} \, .
\end{align}

This probability is bounded away from $1$ if
\begin{align}
K > \frac{32}{\epsilon^2} \frac{1}{M c_\mathrm{min}}  (\log{2d})^2 \, , 
\end{align}
and 
\begin{align}
K > 128\gamma \left[ \frac{2}{\epsilon^3} \frac{d}{M} \log{\left(\frac{20}{\epsilon c_\mathrm{min}} \right)} 
                                     + \frac{1}{4 \epsilon^2} \log{M} \right] \, .
\end{align}


\section{Estimating the factor $\gamma$}\label{Sec:speed}

The goal of this Appendix is to estimate the factor $\gamma$ that determines the secret key consumption rate. 
The objective is therefore to evaluate the first and second moments of the random variable
\begin{equation}\label{defX}
X = | \langle \phi | U | \psi_j \rangle |^2 \, ,
\end{equation}
where $\phi$ restricted to be a vector in the single-occupancy subspace $\mathcal{H}_1$, which is our code space.
A generic state can be written as 
\begin{equation}
|\phi\rangle = \sum_{q,t} \alpha_{q,t} |\phi_{q,t}\rangle \, .
\end{equation}
We can apply the Cauchy-Schwarz inequality as shown in Section \ref{Sec:prel}. This yields (see Eq.~\eqref{gamma1}):
\begin{align}
\gamma 
       & \leq 2 \max_q \frac{ \mathbb{E}_U[ |\langle \phi_{q,t} | U | \psi \rangle |^4 ] }{ \mathbb{E}_U[ | \langle \phi_{q,t} |  U | \psi \rangle |^2 ]^2}  \, .
\label{eq:gamma_q}
\end{align}
By symmetry, the quantities 
\begin{align}
\gamma_q := \frac{ \mathbb{E}_U[ |\langle \phi_{q,t} | U | \psi \rangle |^4 ] }{ \mathbb{E}_U[ | \langle \phi_{q,t} |  U | \psi \rangle |^2 ]^2}
\end{align}
do depend on $q$ but not on the particular vector $\phi_q$ in the subspace $\mathcal{H}_q$, nor on the code word $\psi$.
Therefore for each $q$, $\gamma_q$ can be computed numerically and in turn obtain an estimate for the upper bound on the speed of convergence
\begin{align}
\gamma \leq 2 \max_q \gamma_q \, .
\end{align}

Following the notation from Ref.~\cite{scheel2004permanents},
 let $\mathbf\Lambda[k_1,...,k_m|l_1,...l_m]$ be the $m \times m$ matrix whose matrix elements
 are those of the original matrix  $\mathbf\Lambda$ with row indices $k_1,...,k_m$, and
 column indices $l_1,...l_m$. 
 \begin{align}
 \mathbf \Lambda[k_1,k_2,k_3|l_1,l_2,l_3] = 
 \begin{pmatrix}
 \Lambda_{k_1 l_1} & \Lambda_{k_1 l_2} &  \Lambda_{k_1 l_3} \\
 \Lambda_{k_2 l_1} & \Lambda_{k_2 l_2} &  \Lambda_{k_2 l_3} \\
 \Lambda_{k_3 l_1} & \Lambda_{k_3 l_2} &  \Lambda_{k_3 l_3} 
 \end{pmatrix}
 \end{align}

The object $\mathbf \Lambda[1^{i1},2^{i2},..|1^{j1}, 2^{j2}...]$ denotes a matrix whose entries are taken from the matrix $\mathbf \Lambda$,
and whose row index $l$ occurs $i_l$ times, and whose column index $k$ occurs $j_k$ times.
 For example 
 \begin{align}
 \mathbf \Lambda[1^1, 2^1, 3^1| 1^0,2^2,3^1] &=
 \begin{pmatrix}
 \Lambda_{12} & \Lambda_{12} &  \Lambda_{13} \\
 \Lambda_{22} & \Lambda_{22} &  \Lambda_{23} \\
 \Lambda_{32} & \Lambda_{32} &  \Lambda_{33} 
 \end{pmatrix} \, , \\
 \braket{i_1,i_2,...i_m|\mathbf \Lambda| j_1,j_2,...j_m} &= 
 \left( \prod_l i_l  \right)^{-1/2}  \left( \prod_k j_k  \right)^{-1/2} \text{Perm} \mathbf 
 \Lambda[1^{i_1}, 2^{i_2}...,N^{i_M}|1^{j_1}, 2^{j_2},..., N^{j_M}] \, .
 \label{eq:perm}
\end{align} 
Using Eq.~\eqref{eq:perm}, we can calculate Eq.\ \eqref{eq:gamma_q} for a particular photon occupancy pattern. 

\color{black}
We numerically compute $\gamma_q$ for different photon patterns for $n$ between $2$ and $8$, examples are given in Table \ref{tab:small} and \ref{Table:gamma}. Note that the number of configurations to search over grows exponentially with $n$, and thus the search becomes infeasible with high $n$. 
The calculations were performed in Python by computing the permanents of $n\times n$ submatrices of the $m\times m$ unitaries generated from the Haar measure. The expectation value is taken by averaging over $\sim 10^6$ runs.
We observe that the highest value of $\gamma_q$ is achieved when all the photons populate only one mode. To make the calculation feasible, we conjecture (Conjecture \ref{con:gamma}) that this is also true for higher $n$; in this case, the computation can be performed much more efficiently because the submatrices have repeated rows. 
This conjecture has been used to produce the plots in Fig.\ \ref{Fig:rates}.
We repeat the calculation for $n=9$ to $13$, where the results are shown in Table \ref{table:conj}.

\color{black}


We now consider the regime of $m \gg n^2$ in which we can neglect photon bunching.
Therefore, we compute the first and second moments of the random variable
\begin{equation}
X' = | \langle \psi_{j'} | U | \psi_j \rangle |^2 \, .
\end{equation}
This is a little less general than (\ref{defX}) because $\psi_{j'}$ is not a generic vector
in $\mathcal{H}_n^m$.
In fact $\psi_j$ and $\psi_{j'}$ identify two sets of modes, with labels 
$(i_1,i_2, \dots i_n)$ and $(i'_1,i'_2, \dots i'_n)$, respectively. This corresponds to photon-counting on
the modes, which as we know, maps onto
$n \times n$ sub-matrix $A^{(jj')}$ of the unitary matrix $U$:
\begin{equation}
A_{hk}^{(jj')} := U_{i_h i'_k } \, .
\label{eq:x}
\end{equation} 
The random variable $X'$ is the modulus square of the permanent of $A^{(jj')}$:
\begin{equation}
X' = | \langle \psi_{j'} | U | \psi_j \rangle |^2 = \left| \sum_\pi \prod_{h=1}^n A_{h \pi(h)}^{(jj')} \right|^2 \, ,
\end{equation}
where the sum is over all the permutations $\pi$.

To further explore the statistical properties of the permanent, it is useful to recall
that a given entry of a random $m \times m$ unitary is itself distributed approximately as an complex Gaussian variable with zero mean and variance $1/m$. If instead we consider a submatrix of size $n \times n$ the entries
are with good approximation independent Gaussian variables as long as $n \ll m$ \cite{aaronson2011computational}.
This means that the entries $A_{hk}^{(jj')}$ are {approximately} distributed as  $n^2$ i.i.d.\ complex Gaussian variables
with zero mean and variance $1/m$. Using this fact we can compute the first and second moments of $X'$.

\begin{align}
X' = \left(\sum_{\tau} \prod_{j=1}^n a^*_{j \tau(j)}    \right) \times \left(\sum_{\sigma} \prod_{i=1}^n a_{i \sigma(i)}    \right) 
  = \sum_{\sigma,\tau} \prod_{i,j=1}^n a^*_{j\tau(j)} a_{i \sigma(i)} 
 = \frac{n!}{m^n},
\end{align}
since the non-zero terms are given by $i=j,\tau=\sigma$.

From Lemma 56 of Ref.~\cite{aaronson2011computational}, the fourth moment of the permanent can be computed as
\begin{align}
\mathbb{E}_U[X'^2]= \mathbb{E}_U[\text{Perm}[A]^2\text{Perm}[A^*] ^2] = \frac{n! (n+1)!}{m^{2n}} \, .
\end{align}
In conclusion, we have obtained
\begin{align}
\frac{ \mathbb{E}_U[{X'}^2] }{ \mathbb{E}_U[{X'}]^2 } = n+1 \, .
\end{align}
From which it follows,
\begin{align}
\gamma \leq 2(n+1) \, .
\end{align}


\footnotesize

\begin{table}
\begin{tabular}{ |c|  c| c|  }
\hline
$(m,n)$  & Photon pattern &  $2\gamma_q$      \\ 
\hline
(20,4)  & $ (1, 1, 1, 1,0,\dots) $   &  5.44 \\
& $ (1, 3,0,\dots) $   &  8.91 \\
& $ (1, 1, 2,0,\dots) $   &  6.60 \\
& $ (2, 2,0,\dots) $   &  8.38 \\
& $ (4,0,\dots) $   &  13.31 \\
\hline
(20,5)
& $ (1, 1, 1, 1, 1,0,\dots) $   &  5.45 \\
& $ (1, 1, 3,0,\dots) $   &  8.44 \\
& $ (1, 4,0,\dots) $   &  12.13 \\
& $ (1, 2, 2,0,\dots) $   &  7.80 \\
& $ (2, 3,0,\dots) $   &  10.50 \\
& $ (1, 1, 1, 2,0,\dots) $   &  6.50 \\
& $ (5,0,\dots) $   &  16.93 \\
\hline
(20,6) 
& $ (1, 1, 1, 1, 1, 1,0,\dots) $   &  5.34 \\
& $ (1, 1, 2, 2,0,\dots) $   &  7.26 \\
& $ (3, 3,0,\dots) $   &  12.86 \\
& $ (2, 2, 2,0,\dots) $   &  8.72 \\
& $ (1, 1, 1, 3,0,\dots) $   &  7.77 \\
& $ (1, 5,0,\dots) $   &  15.89 \\
& $ (1, 1, 1, 1, 2,0,\dots) $   &  6.16 \\
& $ (2, 4,0,\dots) $   &  13.42 \\
& $ (1, 1, 4,0,\dots) $   &  10.50 \\
& $ (1, 2, 3,0,\dots) $   &  9.50 \\
& $ (6,0,\dots) $   &  26.34 \\
\hline
(20,8) & $ (1, 1, 6,0,\dots) $   &  15.81 \\
    & $ (1, 1, 1, 2, 3,0,\dots) $   &  7.30 \\
    & $ (4, 4,0,\dots) $   &  18.04 \\
    & $ (2, 2, 4,0,\dots) $   &  11.11 \\
    & $ (1, 1, 1, 1, 2, 2,0,\dots) $   &  5.98 \\
    & $ (2, 2, 2, 2,0,\dots) $   &  7.95 \\
    & $ (2, 3, 3,0,\dots) $   &  10.91 \\
    & $ (1, 1, 1, 5,0,\dots) $   &  11.03 \\
    & $ (2, 6,0,\dots) $   &  19.18 \\
    & $ (1, 1, 3, 3,0,\dots) $   &  9.09 \\
    & $ (1, 1, 1, 1, 1, 1, 1, 1,0,\dots) $   &  4.92 \\
    & $ (1, 7,0,\dots) $   &  24.34 \\
    & $ (8,0,\dots) $   &  34.86 \\ 
\hline
\end{tabular}
\begin{tabular}{ |c|  c| c| }
\hline
$(m,n)$  & Photon pattern &  $2\gamma_q$     \\ 
\hline
(30,4)  
& $ (1, 1, 1, 1,0,\dots) $   &  5.92 \\
& $ (1, 3,0,\dots) $   &  9.97 \\
& $ (1, 1, 2,0,\dots) $   &  7.28 \\
& $ (2, 2,0,\dots) $   &  9.47 \\
& $ (4,0,\dots) $   &  15.07 \\
\hline
(30,5)  & $ (1, 1, 1, 1, 1,0,\dots) $   &  6.12 \\
        & $ (1, 1, 3,0,\dots) $   &  10.10 \\
        & $ (1, 4,0,\dots) $   &  14.63 \\
        & $ (1, 2, 2,0,\dots) $   &  9.30 \\
        & $ (2, 3,0,\dots) $   &  12.88 \\
        & $ (1, 1, 1, 2,0,\dots) $   &  7.44 \\
        & $ (5,0,\dots) $   &  19.12 \\
\hline
(30,6) & $ (1, 1, 1, 1, 1, 1,0,\dots) $   &  6.11 \\
& $ (1, 1, 2, 2,0,\dots) $   &  9.03 \\
& $ (3, 3,0,\dots) $   &  16.74 \\
& $ (2, 2, 2,0,\dots) $   &  11.19 \\
& $ (1, 5,0,\dots) $   &  20.82 \\
& $ (1, 1, 1, 1, 2,0,\dots) $   &  7.29 \\
& $ (2, 4,0,\dots) $   &  17.44 \\
& $ (1, 1, 4,0,\dots) $   &  13.98 \\
& $ (1, 1, 1, 3,0,\dots) $   &  9.72 \\
& $ (1, 2, 3,0,\dots) $   &  11.99 \\
& $ (6,0,\dots) $   &  33.20 \\
\hline
          \hline
(30,8)  & $(1, 1, 6,0,\dots) $   & 23.70\\
        & $(1, 1, 1, 2, 3,0,\dots) $   & 9.63 \\
        & $(4, 4,0,\dots) $   & 27.56\\
        & $(2, 2, 4,0,\dots) $   & 17.58\\
        & $(1, 1, 1, 1, 2, 2,0,\dots) $   & 7.63\\
        & $(2, 2, 2, 2,0,\dots) $   & 10.10\\
        & $(2, 3, 3,0,\dots) $   & 17.84 \\
        & $(1, 1, 1, 5,0,\dots) $   & 15.70\\
        & $(2, 6,0,\dots) $   & 35.89\\
        & $(1, 1, 3, 3,0,\dots) $   & 12.80\\
        & $(1, 1, 1, 1, 1, 1, 1, 1,\dots) $   & 5.82 \\
        & $(1, 7,0,\dots) $   & 34.43\\
        & $(8,0,\dots) $   & 61.69\\
\hline
\end{tabular}
\end{table}
\begin{table} \centering
\begin{tabular}{ |c|  c| c| c| }
\hline
$(m,n)$  & Photon pattern &  $2\gamma_q$       \\ 
\hline
(40,4)  & $ (1, 1, 1, 1,0,\dots) $   &  6.09 \\
        & $ (1, 3,0,\dots) $    &  10.60 \\
        & $ (1, 1, 2,0,\dots) $   &  7.59 \\
        & $ (2, 2,0,\dots) $   &  9.81 \\
        & $ (4,0,\dots) $   &  15.91 \\
\hline
(40,5)  & $ (1, 1, 1, 1, 1,0,\dots) $   &  6.51 \\
        & $ (1, 1, 3,0,\dots) $   &  11.01 \\
        & $ (1, 4,0,\dots) $   &  16.08 \\
        & $ (1, 2, 2,0,\dots) $   &  10.02 \\
        & $ (2, 3,0,\dots) $   &  13.93 \\
        & $ (1, 1, 1, 2,0,\dots) $   &  8.050 \\
        & $ (5,0,\dots) $   &  20.18 \\
\hline
(40,6)  & $ (1, 1, 1, 1, 1, 1,0,\dots) $   &  6.68 \\
        & $ (1, 1, 2, 2,0,\dots) $   &  10.03 \\
        & $ (3, 3,0,\dots) $   &  19.82 \\
        & $ (2, 2, 2,0,\dots) $   &  13.04 \\
        & $ (1, 5,0,\dots) $   &  23.15 \\
        & $ (1, 1, 1, 1, 2,0,\dots) $   &  8.16 \\
        & $ (2, 4,0,\dots) $   &  20.40 \\
        & $ (1, 1, 4,0,\dots) $   &  15.57 \\
        & $ (1, 1, 1, 3,0,\dots) $   &  10.91 \\
        & $ (1, 2, 3,0,\dots) $   &  13.88 \\
        & $ (6,0,\dots) $   &  35.95 \\
          \hline
(40,8)  & $ (1, 1, 6,0,\dots) $   &  32.65 \\
        & $ (1, 1, 1, 2, 3,0,\dots) $   &  12.24 \\
        & $ (4, 4,0,\dots) $   &  47.54 \\
        & $ (2, 2, 4,0,\dots) $   &  23.89 \\
        & $ (1, 1, 1, 1, 2, 2,0,\dots) $   &  9.09 \\
        & $ (2, 2, 2, 2,0,\dots) $   &  14.25 \\
        & $ (2, 3, 3,0,\dots) $   &  20.99 \\
        & $ (1, 1, 1, 5,0,\dots) $   &  21.25 \\
        & $ (2, 6,0,\dots) $   &  36.11 \\
        & $ (1, 1, 3, 3,0,\dots) $   &  17.05 \\
        & $ (1, 1, 1, 1, 1, 1, 1, 1,0,\dots) $   &  6.484 \\
        & $ (1, 7,0,\dots) $   &  56.08 \\
        & $ (8,0,\dots) $   &  67.49 \\
\hline
\end{tabular}
\caption{The table shows numerically computed values of $\gamma_q$ for different number of photons ($n$), modes ($m$), and photon occupancy patterns. }\label{Table:gamma}
\end{table}

\begin{table}
\begin{tabular}{ |c|  c| c| c|}
\hline
$(m,n)$  & Photon pattern &  $2\gamma_q$     \\ 
\hline
\hline
 $(20,9)$ &  $ (9,0,...,0) $   &40.84   \\
$(20,10)$ &  $ (10,0,...,0) $    &53.87\\
\hline \hline
$(30,9)$  &  $ (9,0,...,0) $   &73.45  \\
(30,10)   &  $ (10,0,...,0) $   &111.5\\
(30,11)   &  $ (11,0,...,0) $   &124.6\\
(30,12)   & $ (12,0,...,0) $   &164.7\\
(30,13)   & $ (13,0,...,0) $   &201.4\\
\hline
(40,9)& $ (9,0,...,0) $   & 114.5  \\
(40,10)& $ (10,0,...,0) $   & 161.2 \\
(40,11)& $ (11,0,...,0) $   & 207.2\\
(40,12) & $ (12,0,...,0) $   & 259.7\\
(40,13) & $ (13,0,...,0) $   &422.1\\
\hline
\end{tabular}
\begin{tabular}{ |c|  c| c| c|}
\hline
$(m,n)$  & Photon pattern &  $2\gamma_q$     \\ 
\hline
\hline
(60,4)& $ (4,0,...,0) $   & 16.63 \\
(60,6)& $ (6,0,...,0) $   & 43.59\\
(60,8)& $ (8,0,...,0) $   & 112.6\\
(60,10) & $ (10,0,...,0)$ & 230.2 \\
(60,12) & $ (12,0,...,0)$ & 500.4\\
(60,14) &  $ (14,0,...,0)$ & 722.7 \\
(60,16) & $ (16,0,...,0) $ &1877\\
(60,18) & $ (18,0,...,0) $        & $2.526 \times 10^4$ \\
\hline
\end{tabular}
\caption{\label{table:conj} Values of $\gamma_q$ for the completely bunched photon configuration.}
\end{table}

\clearpage

\section{Mutual information for a pure loss channel}\label{Sec:loss}

For a lossy channel with transmissivity $\eta$, the mutual information 
$I(X;Y|K)$ between Alice and Bob (given that Bob knows the secret key) can be computed explicitly. 
We assume that Bob measures by photo-detection.
First, we note that, since the key is independent on the both $X$ and $Y$, we have $I(X;Y|K) = I(X;Y)$.

If Alice sends one particular code words $\psi_j$ containing $n$ photons, 
Bob will get $k$ photons with probability $p(k|j) = \eta^k (1-\eta)^{n-k}$. 
This is uniquely identified if Bob measures by photo-detection by $k$ detection events.
There exists $N_k = {n \choose k}$ possible measurement outputs of this kind.
Therefore the conditional entropy is
\begin{align}
H(B|X) & = - \sum_j p(j) \sum_k N_k \, p(k|j) \log{p(k|j)} \\
& = - \sum_j p(j)\sum_{k=0}^n {n \choose k} \eta^{k} (1-\eta)^{n-k} \log{\left[ \eta^{k} (1-\eta)^{n-k} \right]} \\
& = - \sum_{k=0}^n {n \choose k} \eta^{k} (1-\eta)^{n-k} \log{\left[ \eta^{k} (1-\eta)^{n-k} \right]} \, ,
\end{align}
where $p(j)$ is the probability of code words $\psi_j$.

Now consider that Bob obtains a certain combination of $k$ detection events over $m$ modes.
This output is compatible with $M_k = {m-k \choose n-k}$ input code words sent by Alice.
As the total number of code words is $M = {m \choose n}$, the probability of obtaining a given
combination of $k$ detections is 
\begin{align}
p(k) & = \frac{M_k}{M} \, p(k|j) 
     = \frac{{ m-k\choose n-k}}{{ m\choose n}} \, \eta^{k} (1-\eta)^{n-k} \, .
\end{align}
Note that the total number of possible outputs is $N'_k = { m \choose k}$, therefore we have
\begin{align}
H(B) & = - \sum_{k=0}^n N'_k p(k) \log{p(k)} \\
& = - \sum_{k=0}^n { m \choose k} \frac{{ m-k\choose n-k}}{{ m\choose n}} \, \eta^{k} (1-\eta)^{n-k} \log{\left[ \frac{{ m-k\choose n-k}}{{ m\choose n}} \, \eta^{k} (1-\eta)^{n-k} \right]} \\
& = - \sum_{k=0}^n {n \choose k} \, \eta^{k} (1-\eta)^{n-k} \log{\left[ \frac{{ m-k\choose n-k}}{{ m\choose n}} \, \eta^{k} (1-\eta)^{n-k} \right]} \, .
\end{align}

Finally we obtain:
\begin{align}
I(X;BK) & = - \sum_{k=0}^n {n \choose k} \eta^{k} (1-\eta)^{n-k} \log{\frac{{ m-k\choose n-k}}{{ m\choose n}}} \\
& = \log{{ m\choose n}} - \sum_{k=0}^n {n \choose k} \eta^{k} (1-\eta)^{n-k} \log{{ m-k\choose n-k}} \, .
\label{eq:iabloss}
\end{align}

\end{document}